\documentclass[letterpaper,twocolumn,10pt]{article}
\usepackage{usenix}
\usepackage[compact]{titlesec}

\usepackage{tikz}
\usepackage{amsmath}
\usepackage{paralist}

\usepackage{txfonts}

\usepackage{amsmath,amsfonts,bm}

\def\eqref#1{equation~\ref{#1}}

\def\1{\bm{1}}

\def\vtheta{{\bm{\theta}}}

\DeclareMathAlphabet{\mathsfit}{\encodingdefault}{\sfdefault}{m}{sl}
\SetMathAlphabet{\mathsfit}{bold}{\encodingdefault}{\sfdefault}{bx}{n}

\DeclareMathOperator*{\argmin}{arg\,min}

\usepackage{comment}

\usepackage{amssymb,pifont}
\usepackage{multirow}
\usepackage{listings}[language=Python, caption=Python example]
\usepackage{footmisc}

\usepackage{diagbox}
\usepackage{slashbox}
\usepackage{etoolbox}
\usepackage{xspace}

\usepackage{booktabs}
\usepackage{threeparttable}
\usepackage[bb=boondox]{mathalfa}
\usepackage{array}
\usepackage{cleveref}

\usepackage{algorithm}
\usepackage{algorithmic}

\usepackage{float}

\usepackage{colortbl}
\definecolor{Gray}{gray}{0.92}
\definecolor{LightCyan}{rgb}{0.92,1,1} 
\newcolumntype{a}{>{\columncolor{Gray}}r}
\newcolumntype{b}{>{\columncolor{LightCyan}}r}

\newcommand{\paragraphbe}[1]{\vspace{0.75ex}\noindent{\bf \em #1}\hspace*{.3em}}

\newcommand{\vit}[1]{{\textcolor{purple}{[VS: \textbf{#1}]}}}

\renewcommand{\L}{\mathcal{L}}

\newcommand{\D}{\mathcal{D}}

\newcommand{\Y}{\mathcal{Y}}

\begin{document}

\date{}

\title{\Large \bf Adversarial Illusions in Multi-Modal Embeddings}

\author{
  {\rm Tingwei Zhang\textsuperscript{$\dagger\ast$}} \quad
  {\rm Rishi Jha\textsuperscript{$\dagger\ast$}} \quad 
  {\rm Eugene Bagdasaryan\textsuperscript{$\ddagger$}} \quad 
  {\rm Vitaly Shmatikov\textsuperscript{$\S$}} \\
  \phantom{\thanks{Comparable contributions.}} \\
  {  \textsuperscript{$\dagger$}Cornell University \quad \textsuperscript{$\ddagger$}University of Massachusetts Amherst \quad \textsuperscript{$\S$}Cornell Tech} \\
  {\small \{tingwei, rjha\}@cs.cornell.edu  \quad eugene@cs.umass.edu  \quad shmat@cs.cornell.edu} 
}

\maketitle

\thispagestyle{empty}

\begin{abstract}
Multi-modal embeddings encode texts, images, thermal images, sounds, and videos into a single embedding space, aligning representations across different modalities (\emph{e.g.}, associate an image of a dog with a barking sound).  In this paper, we show that multi-modal embeddings can be vulnerable to an attack we call ``adversarial illusions.''  Given an image or a sound, an adversary can perturb it to make its embedding close to an arbitrary, adversary-chosen input in another modality.

These attacks are cross-modal and targeted: the adversary can align any image or sound with any target of his choice. Adversarial illusions exploit proximity in the embedding space and are thus agnostic to downstream tasks and modalities, enabling a wholesale compromise of current and future tasks, as well as modalities not available to the adversary. Using ImageBind and AudioCLIP embeddings, we demonstrate how adversarially aligned inputs, generated without knowledge of specific downstream tasks, mislead image generation, text generation, zero-shot classification, and audio retrieval.

We investigate transferability of illusions across different embeddings and develop a black-box version of our method that we use to demonstrate the first adversarial alignment attack on Amazon's commercial, proprietary Titan embedding.  Finally, we analyze countermeasures and evasion attacks. 
\end{abstract}

\begin{figure}[!ht]
 \centering \includegraphics[width=1.00\linewidth]{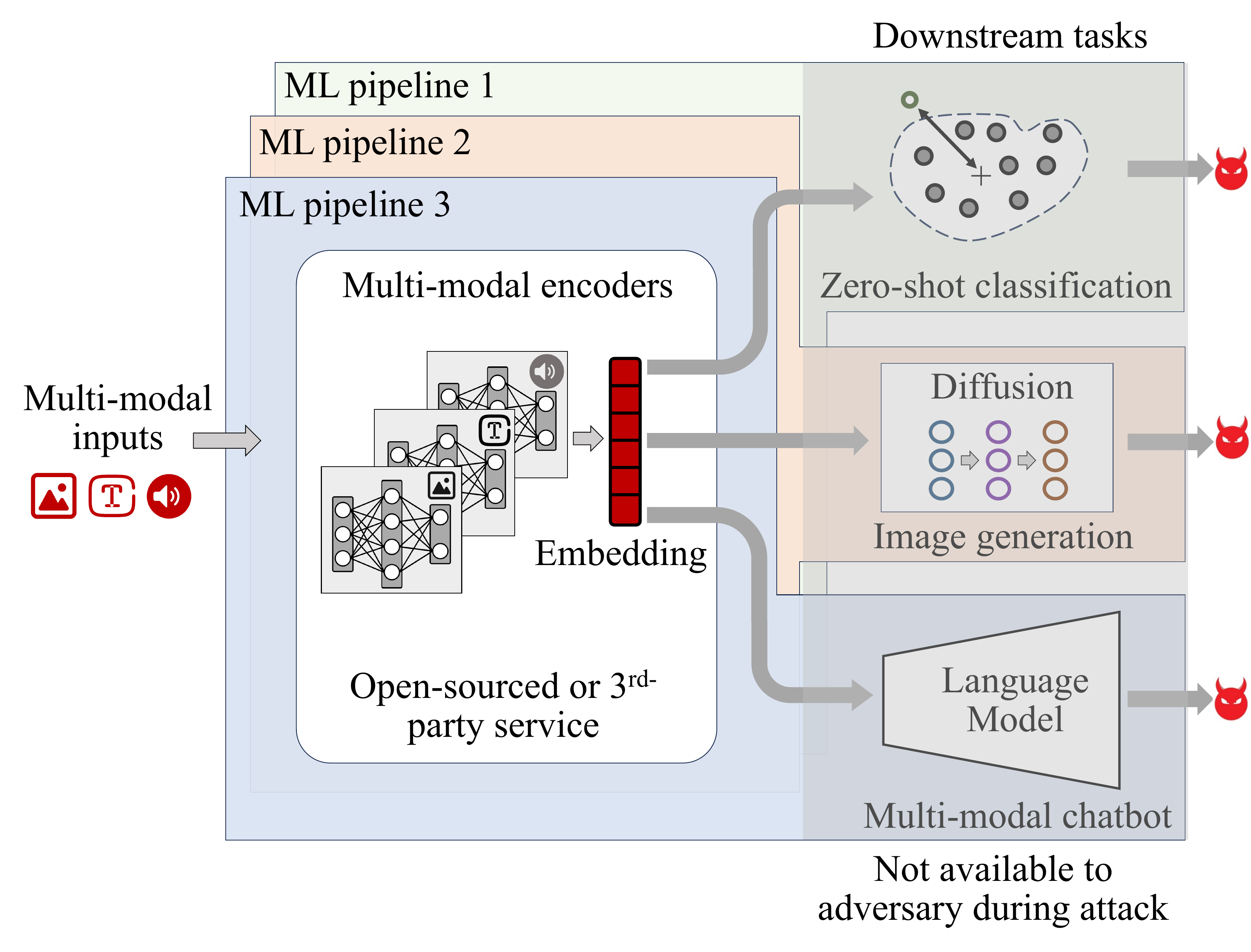}
 \vspace{-2.0ex}
 \caption{\textbf{An adversarial illusion compromises any machine learning (ML) pipeline that relies on the multi-modal encoders.}}
 \label{fig:ml_pipeline}
\end{figure}

\begin{figure*}[tb]
    \centering    \includegraphics[width=1.0\linewidth]{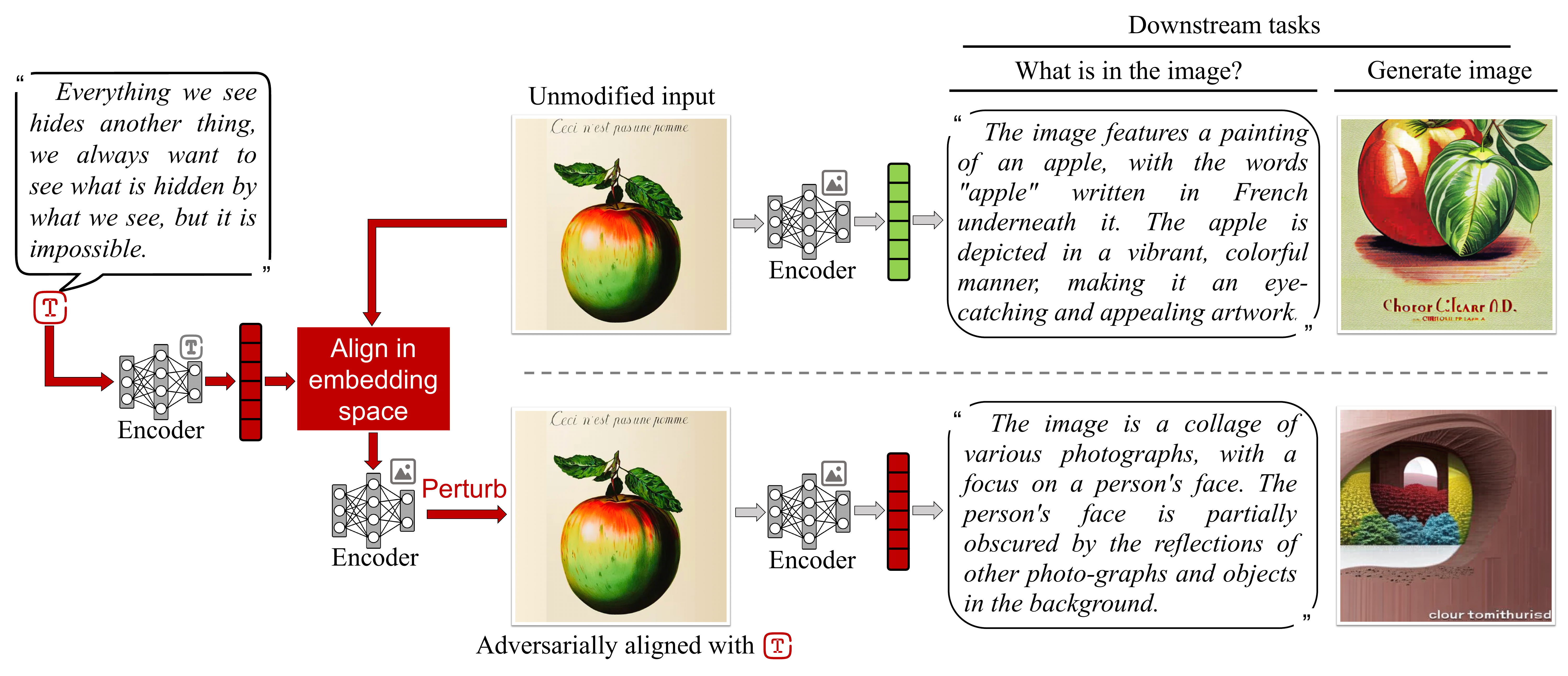}
    \vspace{-2.0ex}
    \caption{\textbf{``Ceci n'est pas une pomme'': an adversarial perturbation
    of an image misleads downstream tasks.}}
    \label{fig:magritte}
\end{figure*}

\vspace{8ex}
\section{Introduction}

Multi-modal encoders such as ImageBind~\cite{girdhar2023imagebind} and AudioCLIP~\cite{guzhov2021audioclip} are a new class of machine learning models that encode inputs from different modalities (\emph{e.g.}, text, audio, conventional and thermal images) into task- and modality-agnostic vectors. Beyond compression, the purpose of multi-modal encoders is to map semantically related inputs into similar vectors so that, for example, the respective embeddings of a dog image and a barking sound are \emph{aligned}, \emph{i.e.}, lie close to each other in the embedding space.  This enables arbitrary combinations of downstream tasks (classification, generation, etc.) and modalities to operate on the same input space.

Modern encoders capture semantically rich spatial relationships between inputs by contrastively training on vast amounts of bi-modal data (\emph{e.g.}, audio recordings and the corresponding captions).  The resulting encoders not only achieve \textit{natural alignment} between these modalities but also exhibit \emph{emergent alignment} between modalities that were never explicitly linked in the training data.  For example, a multi-modal embedding trained on image-text and audio-text pairs may also align semantically related images and sounds even though it was not trained on image-audio pairs.


Cross-modal semantic relationships encoded in the embeddings make them a useful ``backbone'' building block for current and future downstream tasks, without training models for these tasks on multiple modalities.  They also present a new attack surface, which we explore in this paper.

We demonstrate that multi-modal embeddings are vulnerable to a new class of attacks that were not possible with previous embeddings.  By aligning inputs of his choice in a multi-modal embedding, an adversary can compromise output modalities and downstream tasks that are not available or even known at the time of the attack, as shown in Figure~\ref{fig:ml_pipeline}.

\begin{figure}[!ht]
  \centering
  \begin{minipage}[t]{1.0\linewidth}
    \centering    \includegraphics[width=1.00\linewidth]{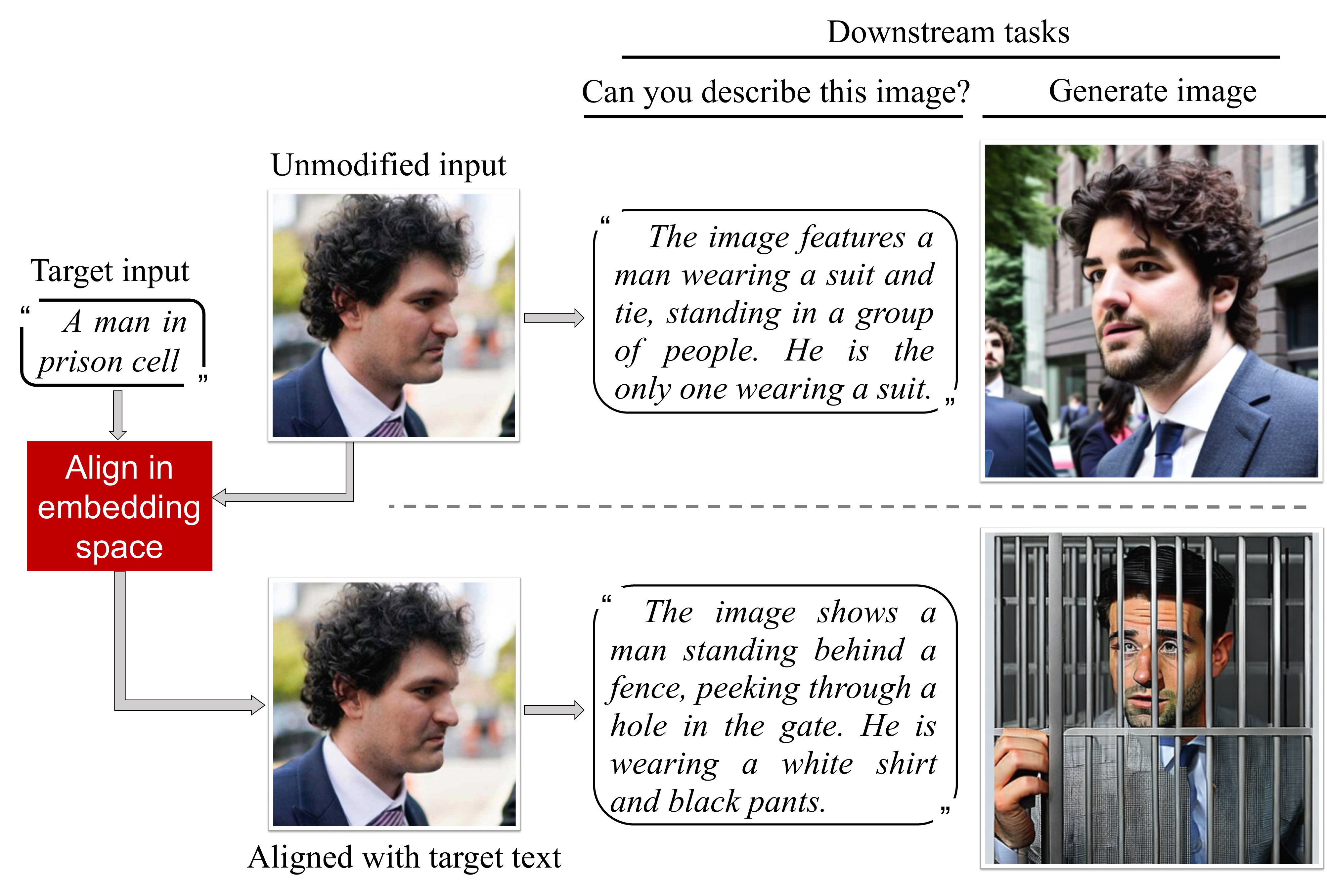}
    \vspace{-3ex}
    \caption{\textbf{``Schadenfreude'': a visual illusion against image and text generation.}}
    \label{fig:person}
  \end{minipage}
  
  \vspace{4.5ex} 
  
  \begin{minipage}[t]{1.0\linewidth}
    \centering    \includegraphics[width=1.0\linewidth]{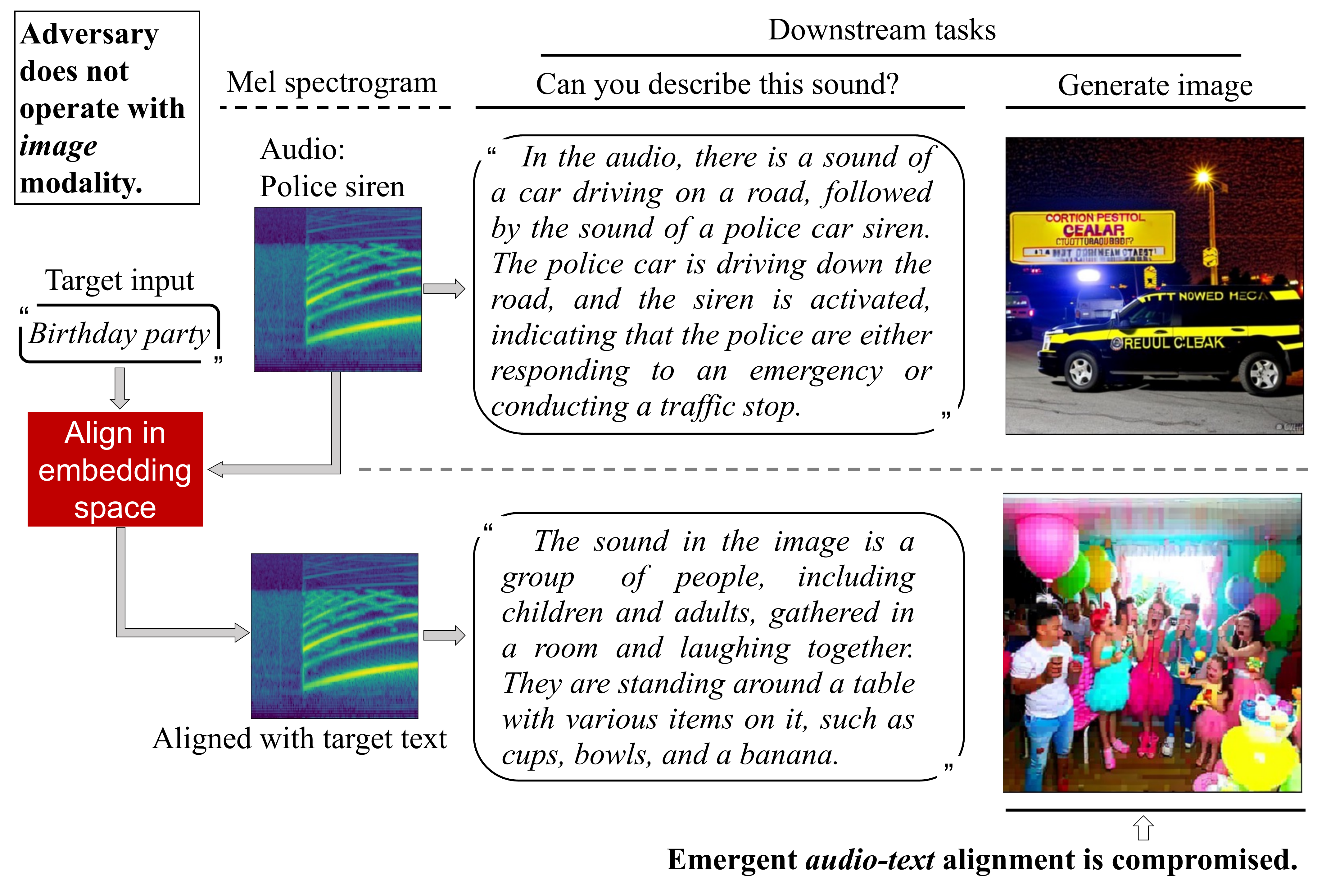}
    \vspace{-3ex}    \caption{\textbf{``Party time'': an audio illusion against image and text generation.}}
    \label{fig:example_siren}
  \end{minipage} 
\end{figure}

Specifically, we show that cross-modal alignment in multi-modal embeddings is highly vulnerable to adversarially generated ``illusions.''  Given an input $x$, we say that a perturbation $x_{\delta}$ is an \emph{illusion} if it appears similar to $x$ to a human but aligns with an adversary-chosen $y$ from another modality in the embedding space. An illusion thus ``misrepresents'' its semantic content to downstream tasks. For example, Figure~\ref{fig:magritte} shows an illusion that aligns an image of Magritte's famous ``This is Not an Apple'' painting with the text of Magritte's quote, ``Everything we see hides another thing.'' \Cref{fig:person,fig:example_siren} show other examples.  In all cases, perturbed images and audio recordings appear similar to their respective original inputs, but downstream tasks\textemdash text and image generation, in these examples\textemdash act on the perturbed inputs as if their semantics matched the adversary-chosen texts.

\paragraphbe{Our contributions.}
First, we demonstrate that tiny adversarial perturbations can be used to incorrectly align inputs from different modalities in the embedding space and thus mislead downstream applications.  We use standard adversarial perturbation techniques for this attack.

The attack is \textbf{targeted}. The adversary has full freedom to choose misaligned inputs and can align any target with any input of his choice from a different modality.  This is also the first attack that exploits emergent alignment to attack downstream output modalities not available to the adversary.  

Second, we show that the attack is \textbf{task-agnostic}.  By design~\cite{girdhar2023imagebind, guzhov2021audioclip}, downstream tasks that use multi-modal embeddings do not consider input modalities and can operate on any vector from the embedding space.  These tasks rely on the alignment between the embeddings of semantically related inputs from different modalities, whether natural or emergent. As we show, this ``organic'' alignment corresponds to relatively weak proximity in the multi-modal embedding space. Whereas previous attacks on multi-modal learning targeted a single modality and did not need to deal with the modality gap~\cite{liang2022mind}, we demonstrate that adversarial alignment can be made significantly closer than any organic alignment regardless of source and target modality. 

Therefore, the adversary can leverage an attack on a multi-modal embedding into a wholesale compromise of current and future downstream tasks (see Figure~\ref{fig:ml_pipeline}), including tasks he is unaware of during the attack. We show that multiple tasks\textemdash including image generation, text generation, audio retrieval, and different types of zero-shot classification (image, thermal image, and audio)\textemdash based on embeddings such as ImageBind, AudioCLIP, OpenCLIP~\cite{cherti2023reproducible}, and Amazon's Titan\footnote{\url{https://aws.amazon.com/bedrock/titan/}} are all misled by cross-modal illusions.  Against zero-shot tasks based on ImageBind and AudioCLIP embeddings, our illusions achieve near-perfect ($>99\%$) attack success rates at standard perturbation bounds.

Third, we analyze \textbf{transferability} of the attack across different encoders, investigate how it is influenced by model architecture, and craft illusions that work against multiple embeddings. In particular, we show that illusions generated using OpenCLIP encoders also achieve $100\%$ and $90\%$ attack success rates against zero-shot image classification on ImageBind and AudioCLIP embeddings.

Fourth, we demonstrate a \textbf{black-box, query-based} attack and show that it is effective against several downstream tasks.  We then combine our query-based and transfer techniques into a hybrid method and use it for the first adversarial alignment attack against Amazon's Titan, a commercial, black-box embedding.  Our hybrid method fools Titan on $42\%$ of our zero-shot image classification examples, even though the embedding is proprietary and completely opaque.

Fifth, we survey several countermeasures and demonstrate how adversarial illusions can evade defenses based on feature distillation (\emph{e.g.}, JPEG compression) and anomaly detection based on consistency of augmentations. Our evasion attacks achieve $88\%$ and $94\%$ success rate against the JPEG defense. Even if multi-modal embeddings were certifiably robust to input perturbations (they are not), we show that certification may not prevent these attacks due to the tradeoff between insufficient robustness and excessive invariance.

The main technical innovations of this work are: (1)
establishing that adversarial alignment in multi-modal embeddings can be made closer than any organic alignment regardless of input and target modality; (2) demonstrating that the scope of this attack is much broader than conventional adversarial examples because it is task-agnostic and affects all models based on the embedding, even those that do not accept the adversary’s target modality as input; and (3) a new hybrid technique for query-based attacks on black-box embedding.

\paragraphbe{Significance of the attacks.}
Many modern ML-based systems operate on third-party content that is not created or controlled by their users.  For example, generative ML systems analyze and summarize webpages and social media, ML-based personal assistants process emails and instant messages, and retrieval-augmented generation and recommender systems operate on databases of content from many users.  Multi-modal embeddings extend input domains of these systems to new modalities without re-training or fine-tuning.

It is a core security principle that any third-party content is potentially adversarial and, therefore, cannot be trusted.  Vulnerabilities described in this paper enable adversaries to craft multi-modal content that is interpreted differently by humans and ML systems.  The resulting exploits can cause ML systems to retrieve or recommend adversary-manipulated content in response to certain queries, interpret it in a way favorable to the adversary, conceal dangerous or harmful content from ML-based detection tools, or else create an illusion of danger or harm where there is none (\emph{e.g.}, see Fig.~\ref{fig:thermal_generation}).

\vspace{1ex}
To facilitate research on the security of multi-modal embeddings, we released our code and models.\footnote{\url{https://github.com/ebagdasa/adversarial_illusions}}

\section{Background and Related Work}


A machine learning (ML) model takes input $x \in \mathcal{X}$ and produces output $y \in \mathcal{Y}$ that can be, for example, a class label for $x$, a generated image, or generated text.  
Today, standard ML pipelines use an encoder $\theta$ (\emph{e.g.}, CLIP~\cite{clip}) to convert inputs into embedding vectors.
Since training an encoder is expensive, many ML pipelines rely on pre-trained encoders (open-sourced or commercial).  
With a pre-trained encoder, it is sufficient to operate on embeddings, rather than raw inputs, in either a zero-shot fashion, or with a trained model.  Many ML pipelines thus have two stages:
\begin{align*}
 \text{1.}\;\; \theta(x) &= e \;\;\text{ --- embedding of input} \\
 \text{2.}\;\; \phi(e) &= y \;\;\text{ --- downstream task}
\end{align*}

\begin{figure}[!t]
  \centering  \includegraphics[width=1.0\linewidth]{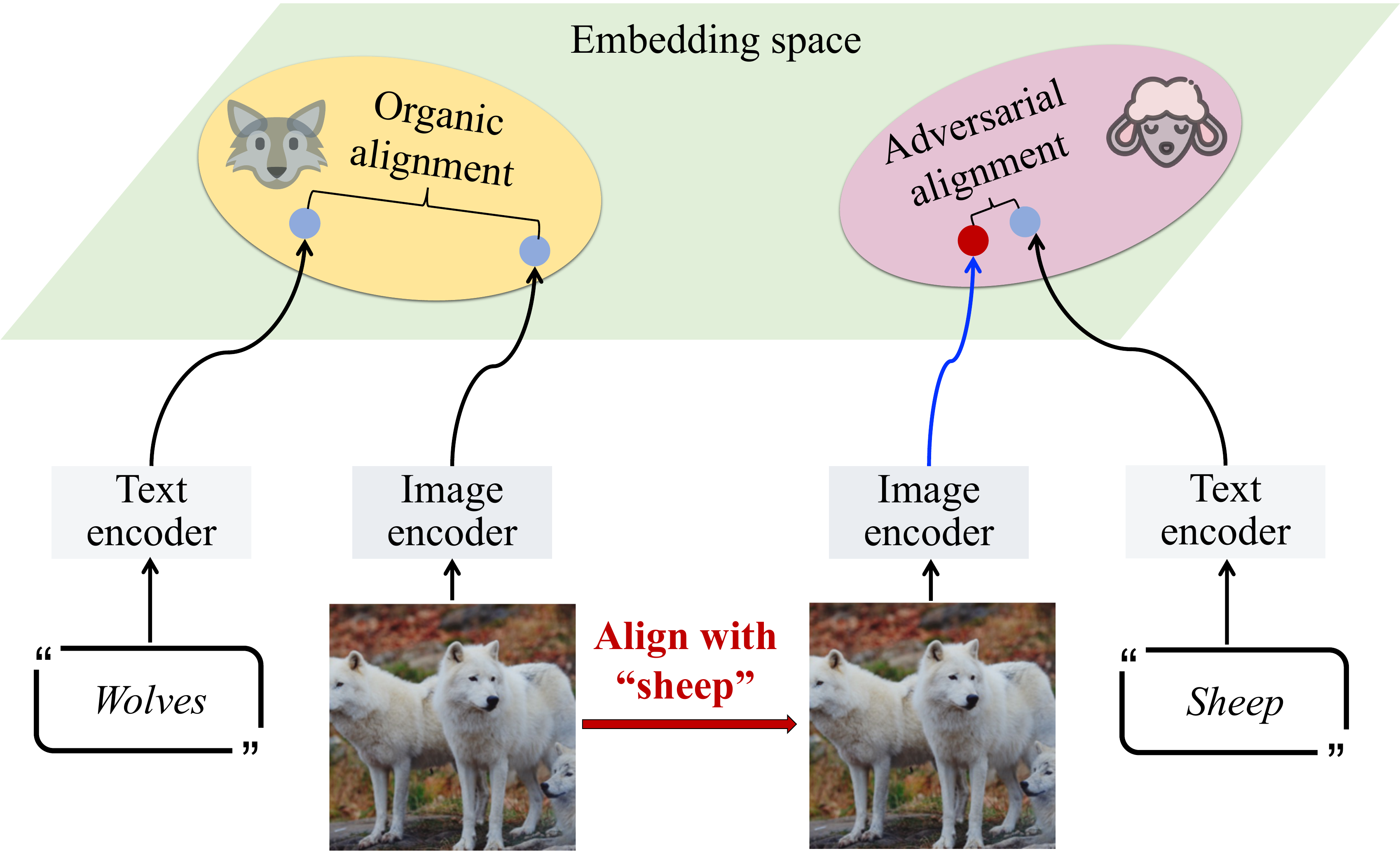}
  \vspace{-2ex}
  \caption{\textbf{Cross-modal, adversarial alignment in the embedding space.}}
  \label{fig:attack}
\end{figure}

\paragraphbe{Multi-modal embeddings.} 
Multi-modal encoders map inputs $x^{(m)}$ from modalities $m \in \mathcal{M}$ into a single embedding space. Each modality $m$ has its own encoder $\theta^{(m)}$.  Encoders are trained on multi-modal tuples $(x^{(m)}, y^{(\bar{m})})$  that are semantically \emph{aligned} (\emph{e.g.}, a picture of wolves and the text ``Wolves,'' as shown in Figure~\ref{fig:attack}).  Encoders are trained using contrastive learning~\cite{oord2018representation} that pushes the representations of aligned inputs closer to each other in the embedding space and away from the representations of semantically different inputs.

Given a bi-modal dataset $\D = (X, Y) \in m \times \bar{m}$, we
define an encoder's \textit{alignment} on $\D$ as the mean cosine similarity between the embeddings of the elements in each $(x, y) \in \D$:
\begin{equation}
\operatorname{align}(\D) = \frac{1}{|\D|}\sum_{(x, y) \in \D} \cos\left(\theta^{(m)}(x), \theta^{(\bar{m})}(y)\right)
\end{equation}

In this paper, we focus on images, sounds, and texts.  As of this writing, ImageBind~\cite{girdhar2023imagebind} and AudioCLIP~\cite{guzhov2021audioclip} are the only open-source multi-modal embeddings that support these three modalities. Multi-modal embeddings exhibit ``emergent'' alignment between modalities. Semantically similar images and sounds (\emph{e.g.}, a picture of a dog and an audio recording of a dog barking) have similar embeddings, even though the training data does not include image-audio tuples.

The distinction between natural (present in the training data) and emergent alignment is important for multi-modal encoders, but attacks in this paper are agnostic to the target modality and the source of its alignment.  We generically define an encoder's \textit{organic} alignment on datasets $\D$ where for all $(x, y)\in \D$, $x$ and $y$ are semantically related, regardless of whether it is learned from explicitly aligned training pairs or ``emerged'' from other modalities.

\paragraphbe{Downstream models.}
Multi-modal embeddings align semantically similar inputs across modalities. Thus, downstream models do not need to be trained on multiple modalities and can operate directly on the embeddings.

\emph{Zero-shot classification} matches the embedding of an input to the closest class embedding (mean embedding of known class representatives). With multi-modal embeddings, zero-shot classification can classify inputs to classes in a different modality, {\em e.g.}, label an audio with an image class. \emph{Retrieval}, similar to zero-classification, matches a query to the closest item in the embedding space. 

\emph{Image generation} takes an embedding and performs conditional generation using a diffusion model. ImageBind's image encoder is initialized from the CLIP visual encoder~\cite{clip}, thus diffusion models that operate on CLIP embeddings (\emph{e.g.}, unCLIP~\cite{ramesh2022hierarchical}) can also operate on ImageBind embeddings. \emph{Text generation} can use multi-modal embeddings as inputs to instruction-following language models, \emph{e.g.}, PandaGPT~\cite{su2023pandagpt}.

\paragraphbe{Adversarial perturbations.}
An adversarial perturbation $\delta$ added to input $x$ causes a model $\phi$ to produce an incorrect output, as in untargeted adversarial examples~\cite{goodfellow2014explaining}, or even a specific output $y_t$ chosen by the adversary~\cite{carlini2018audio}.

Adversarial examples for text target toxicity classification~\cite{hosseini2017deceiving} and reading comprehension~\cite{jia2017adversarial}. Recently, jailbreaking attacks on dialog systems, \emph{i.e.}, chatbots, were demonstrated to bypass internal guardrails on hate speech~\cite{zou2023universal}. All of these attacks target a specific downstream task.

Several recent papers~\cite{yu2022adversarial, kim2020adversarial, zhou2023downstream} deploy adversarial perturbations against encoders that use contrastive learning. These attacks work by increasing contrastive loss and are necessarily \emph{untargeted}: they prevent perturbed inputs from being embedded near the original input, but the adversary has no control over the placement or alignment of perturbed inputs.

In other recent work~\cite{carlini2023aligned, qi2023visual, bagdasaryan2023ab}, adversarial perturbations are used for jailbreaking and prompt injection in multi-modal chatbots (a specific downstream task). In~\cite{shayegani2023plug}, adversarial perturbations against CLIP image embeddings are shown to affect downstream tasks.  The attack is not cross-modal, the adversary has no arbitrary choice of inputs, and perturbations are very large and visible. 

Dong et al.~\cite{dong2023robust} present an untargeted attack on image embeddings and a targeted attack on text descriptions in a specific language model (rather than multi-modal embeddings). The attack causes the model to predict the main object in an image incorrectly.  There is no analysis of how an embedding attack transfers to multiple downstream tasks, modalities beyond images and text, or robustness of defenses.

Zhao et al.~\cite{zhao2023evaluating} present several attacks against visual chatbots, evaluated via the embedding alignment (CLIP Score) between the adversarial image $x_\delta$ and target $y_t$. This is insufficient to show transferability to downstream tasks. For example, to attack a generation model $\phi$ based on embedding $\theta$, it is not enough to show that $\theta(x_\delta)$ and $\theta(y_t)$ are aligned.   The attack succeeds only if $\phi(\theta(x_\delta))$\textemdash \emph{i.e.}, the image \emph{generated} from the embedding, rather than the image used to produce the embedding\textemdash is classified to $y_t$.  Modalities are limited to images and text, transferability across encoders is limited to CLIP as the target, and there is no analysis of defenses, even though adversarial images are foiled by simple countermeasures such as applying JPEG (see Section~\ref{sec:jpeg-defense}).

Our paper demonstrates that adversarial perturbations enable cross-modal, downstream task-agnostic adversarial alignment of arbitrary inputs in multi-modal embeddings. 

\paragraphbe{Other attacks.} 
Previous work considered collisions in NLP models~\cite{song2020adversarial}, but\textemdash unlike this paper\textemdash it targets specific downstream tasks in a single modality (text), and the adversary does not have an arbitrary choice of colliding inputs.

Poisoning and backdoor attacks~\cite{jia2021badencoder, jha2023label} compromise embeddings at \textit{training time}.  By contrast, this paper focuses on \textit{evaluation-time} attacks against ``clean,'' unmodified models.

\section{Threat Model}
\label{sec:threat-model}

In this section, we briefly explain the adversary's goals and capabilities for different attack scenarios. 

\subsection{Adversary's Goals}

The adversary seeks to influence a multi-modal embedding\textemdash and thus any downstream task or, in general, machine learning pipeline that uses the embedding, even if not known to the adversary at the time of the attack\textemdash by crafting a semantically meaningful input we call an \textit{adversarial illusion} so that its embedding is aligned with another, attacker-chosen input in a different modality.  Technically, the adversary's goal is to maximize \textit{adversarial alignment} between the adversary's inputs $x_\delta$ in modality $m$ and (arbitrarily chosen) targets $y_t$ in modality $\bar{m}$\textemdash see Figure~\ref{fig:attack}.

Cross-modal adversarial alignment extends the scope of the attack beyond the modalities supported by a particular ML pipeline. For example, there may not even exist an input in the right modality, {\em e.g.}, there may not be an available image corresponding to the adversary's target text (see an example in Figure~\ref{fig:magritte}).  Furthermore, the attack may target downstream applications that do not accept the adversary's modality as input.  For example, consider an accessibility application that takes images and generates audio describing their contents.  An adversarial alignment attack enables an adversary to use any text of his choice as the target and attack this application with an image aligned with the chosen text\textemdash even though the application does not accept text inputs.

\subsection{Adversary's Capabilities}

We consider four distinct attack settings.  In each setting, the adversary has different levels of access to the target encoders and auxiliary information.

\paragraphbe{White-box.}
Several multi-modal encoders are available as open-sourced code. In this setting, the adversary has access to all of their internal details, including architecture, parameters, and gradients (but not downstream tasks).  This level of access allows the adversary to backpropagate gradients from the loss functions defined in \Cref{sec:methods} directly to adversarial illusions, similar to conventional adversarial examples~\cite{goodfellow2014explaining,madry2018towards}.

\paragraphbe{Transfer.}
If the adversary does not have access to the target encoder (because it is proprietary or closed-source or charges too much for queries), he may use white-box access to other, \textit{surrogate} encoders to create adversarial illusions. The success of the transfer attack depends on the similarity between the surrogate and target models (which may not be known to the adversary in advance) and how well the surrogate attack generalizes~\cite{goodfellow2014explaining,kurakin2016adversarial}.

In \Cref{sec:transfer-obj} and \Cref{sec:transfer-res}, we show that backpropagating gradients from multiple surrogates helps generate illusions that succeed against target encoders to which the adversary does not have access.

\paragraphbe{Query-based.}
Many commercial embeddings are only available as APIs.
In this setting, the adversary has query access to the target encoder but no surrogates.  The adversary has no visibility into the model being queried.


Most query-based attacks rely on estimating the gradients of the target model, require additional auxiliary information, do not work well for targeted attacks, or fail to scale to a large number of queries~\cite{suya2023sok}.  Instead, our attack optimizes over a gradient-free objective (\Cref{sec:BBQ-obj}) to create effective adversarial illusions (\Cref{sec:BBQ-res}). 

\paragraphbe{Hybrid.}
Finally, we investigate a realistic, combined threat model in which the adversary has access to white-box surrogates (\emph{e.g.}, open-source encoders) and query-only access to the target (\emph{e.g.}, commercial embeddings). The adversary still has no prior knowledge about the target but can ``warm-start'' the query-based search with adversarial illusions crafted to transfer across surrogate encoders.

In \Cref{sec:hybrid-obj} and \Cref{sec:hybrid-res}, we show that this strategy, 
starting with white-box access to surrogate encoders, yields a successful query-only attack on Amazon's Titan embeddings.

\vspace{2.5ex}
\section{Crafting Cross-Modal Illusions}
\label{sec:methods}

Let $m, \bar{m} \in \mathcal{M}$ be the modalities supported by a multi-modal encoder $\theta = \{\theta^{(i)}\}_{i \in \mathcal{M}}$.  The adversary's goal
is to generate a perturbation $\delta$ such that the embedding $\theta^{(m)}(x_\delta)$ of $x_{\delta} = x^{(m)} + \delta$ is close to a target $\theta^{(\bar{m})}(y_t^{(\bar{m})})$, by optimizing the following objective:
\begin{equation}
    \label{eq:optimization}  \argmin_{\delta}\left[\L(x_{\delta}, y_t) = \L\left(x^{\left(m\right)}+\delta, y_t^{(\bar{m})};\theta \right) \right]
\end{equation}
In the rest of this section, we define the objective functions $\L$ for each threat model from \Cref{sec:threat-model}.

\subsection{White-Box Attack}
\label{sec:white-box-obj}
Although ImageBind~\cite{girdhar2023imagebind} and AudioCLIP~\cite{guzhov2021audioclip} use dot product during training, modalities that are not naturally aligned (\emph{e.g.}, audio and text in ImageBind) have different normalizations. 
Hence, for our white-box attack, we omit the norms, use cosine similarity, and minimize the following objective:
\begin{equation}
    \L_{\mathrm{WB}}(x_{\delta}, y_t) = 1 - \cos\left(\theta^{(m)}(x_{\delta}), \theta^{(\bar{m})}(y_t)\right)
    \label{eq:whitebox}
\end{equation}

To optimize \Cref{eq:whitebox}, we iteratively update perturbation $\delta$ with Projected Gradient Descent (PGD)~\cite{madry2018towards}. 

\subsection{Transfer Attack}
\label{sec:transfer-obj}

For the transfer setting, we adopt a variation on the ensemble-based approach of~\cite{liu2016delving},  extending our white-box attack to an ensemble of $K$ surrogate models $\Theta = \{\theta_i\}_{i=1}^K$. Each model's weight is parameterized by $\lambda = \left [\lambda_1 \; ... \;\lambda_K \right]$ where $\lambda_i = \frac{1}{K}$. Concretely, we minimize the following objectives:
\begin{equation}
    \label{eq:transfer}
    \L_{\mathrm{T}}(x_{\delta}, y_t) = \textstyle\sum_{\theta \in \Theta} \lambda_i \L_{\mathrm{WB}}(x_{\delta}, y_t; \theta)
\end{equation}

As in the white-box setting, we use a variant of PGD, varying the target model in each iteration of training.

\subsection{Query-Based Attack}
\label{sec:BBQ-obj}

In the black-box setting, the adversary does not have access to the target model's parameters.  We adopt the Square attack~\cite{andriushchenko2020square}, a score-based method that does not rely on local gradient information.  When the adversary's goal is embedding alignment, \emph{i.e.}, cosine distance between the respective embeddings of some input and a target, the attack can be made stronger if the adversary knows non-target labels $\mathcal{Y}nt$ (\emph{e.g.}, in the case of image-text alignment, $\mathcal{Y}nt$ are labels that neither describe the image, nor are targets of the attack).
For $f(x_{\delta}, y_t) = 1 - \L_{WB}(x_{\delta}, y_t)$ (the embedding cosine distance), we update $\delta$ with a randomized search scheme, minimizing
\begin{equation}
 \L_{BB}(x_\delta, y_t) = - f(x_\delta, y_t) + \log\left(\textstyle\sum\nolimits_{y \in \mathcal{Y}nt} e^{f(x_{\delta}, y)} \right)
\end{equation} 

This is a contrastive approach that maximizes proximity to a specific target while minimizing proximity to non-targets.

\subsection{Hybrid Attack}
\label{sec:hybrid-obj}

In the hybrid attack setting, we ``warm-start'' ({\em i.e.}, initialize) a query-based attack with the illusions generated locally by the transfer attack. Even if they do not fully transfer to the target model, we observe that they are still closer in the embedding space to the targeted region than the original input~\cite{suya2020hybrid}.

\begin{figure*}[!ht]
  \begin{minipage}[t]{1.00\linewidth}
    \centering    \includegraphics[width=1.0\linewidth]{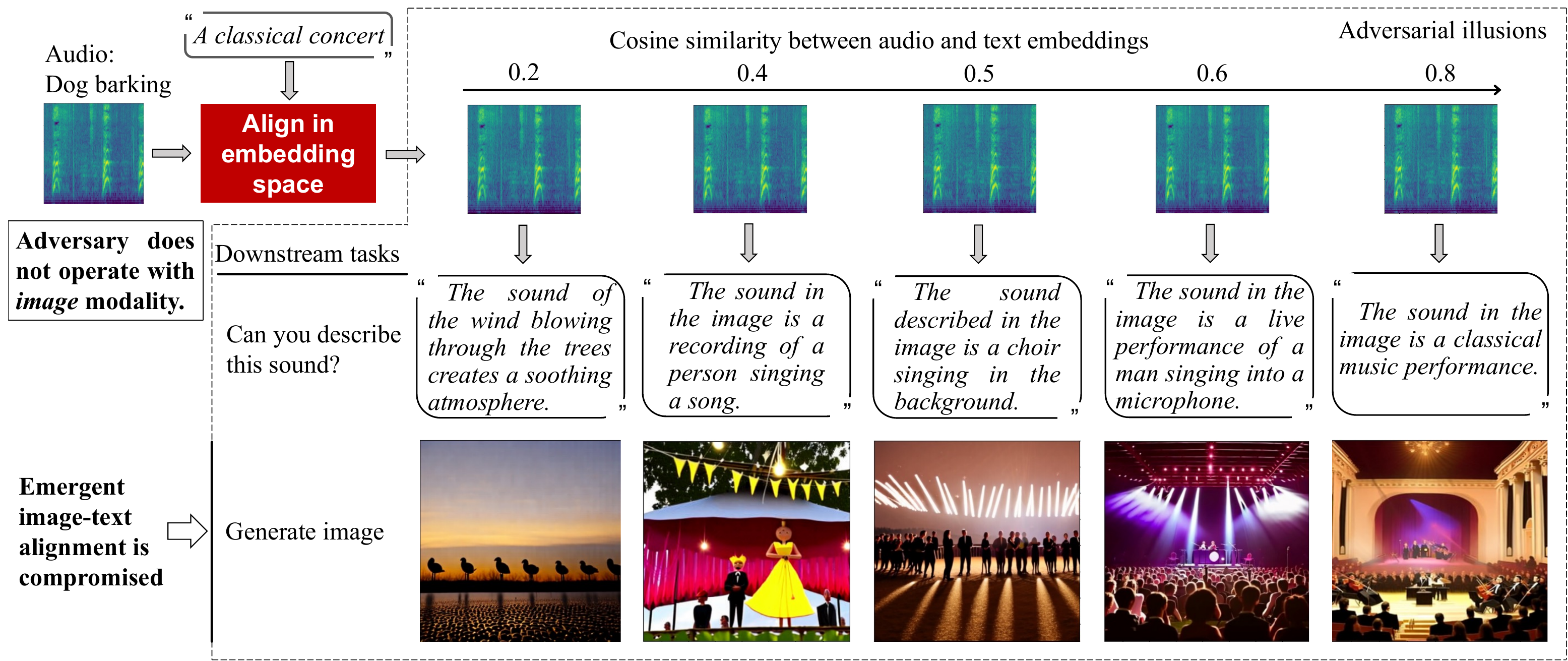}
    \vspace{-1ex}
    \caption{\textbf{``Symphony of Woofs'': similarity between an adversary-chosen input and the resulting illusion.}}
    \label{fig:progress}
  \end{minipage}
  
  \vspace{20pt}

  \begin{minipage}[t]{0.48\linewidth}
    \centering    \includegraphics[width=1.04\linewidth]{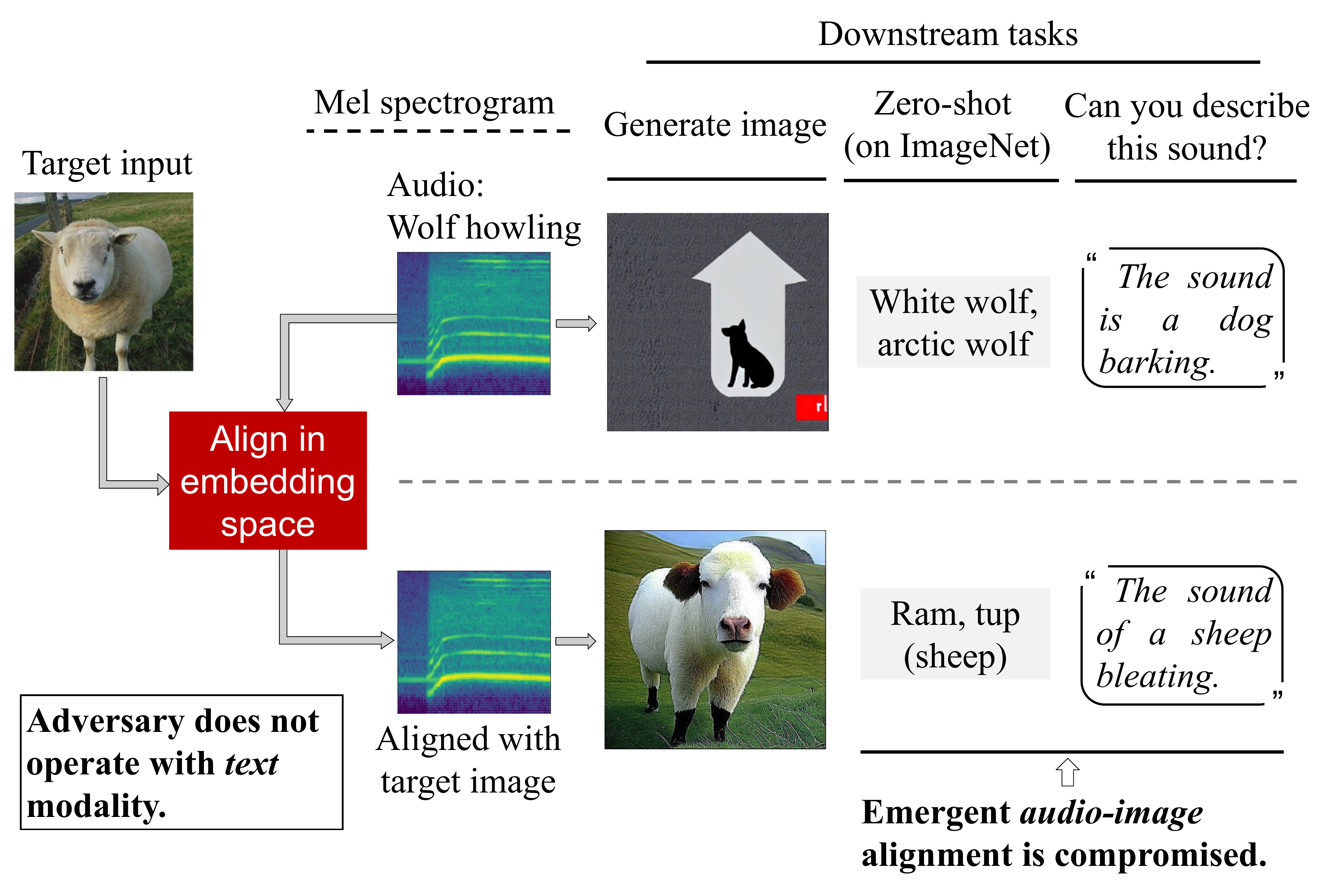}
    \vspace{-12pt}
    \caption{\textbf{``Wolf in sheep's clothing'': an audio illusion
    against image generation, zero-shot classification, and text generation.}}
    \label{fig:zero_shot}
  \end{minipage}
  \qquad 
  \begin{minipage}[t]{0.48\linewidth}
    \centering    \includegraphics[width=0.96\linewidth]{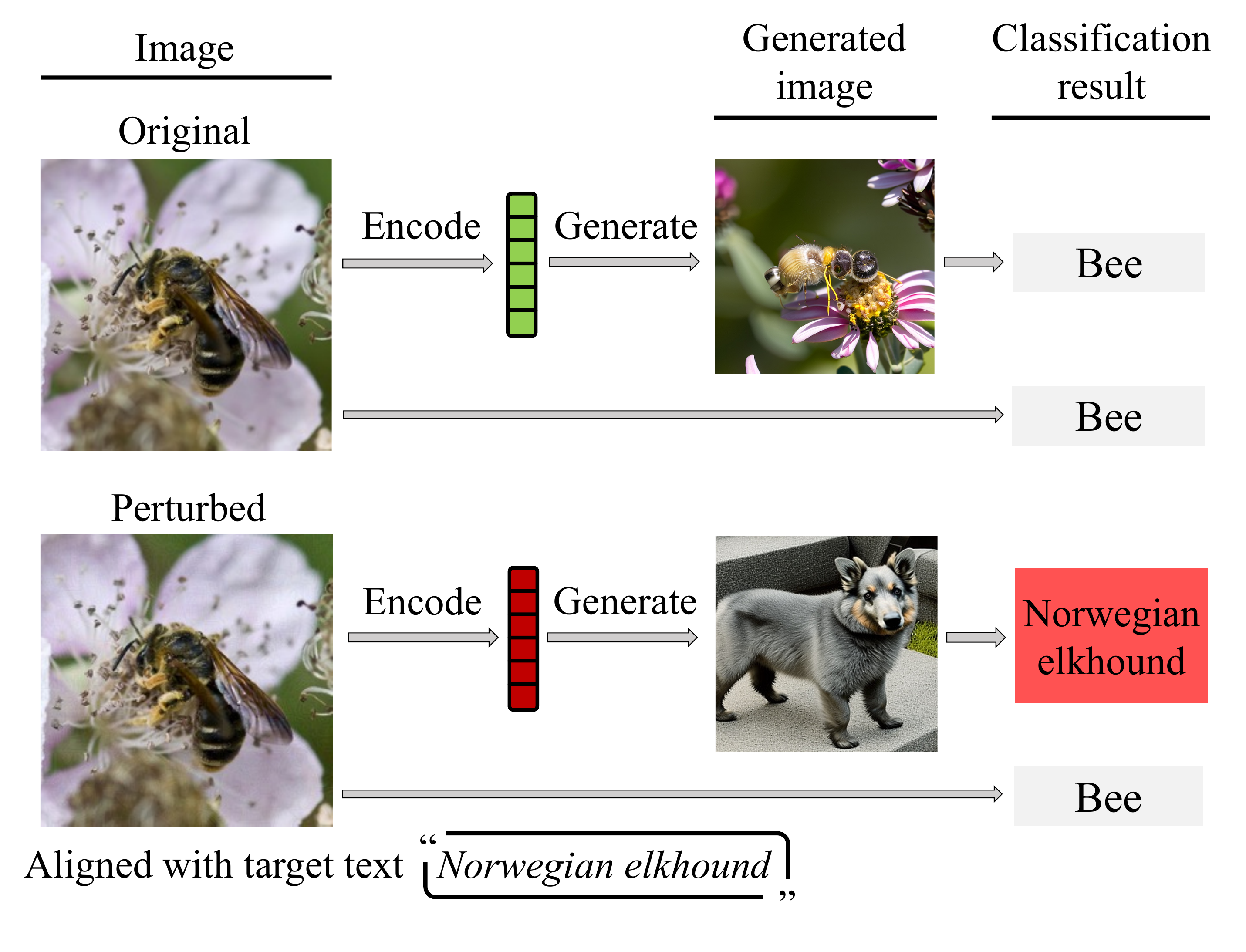}
    \caption{\textbf{``Beewulf'': the original and perturbed images, images generated from their respective embeddings, and their classification.}}
    \label{fig:generative}
  \end{minipage} 
\end{figure*}

\begin{figure}[!t]
  \centering  \includegraphics[width=1.0\linewidth]{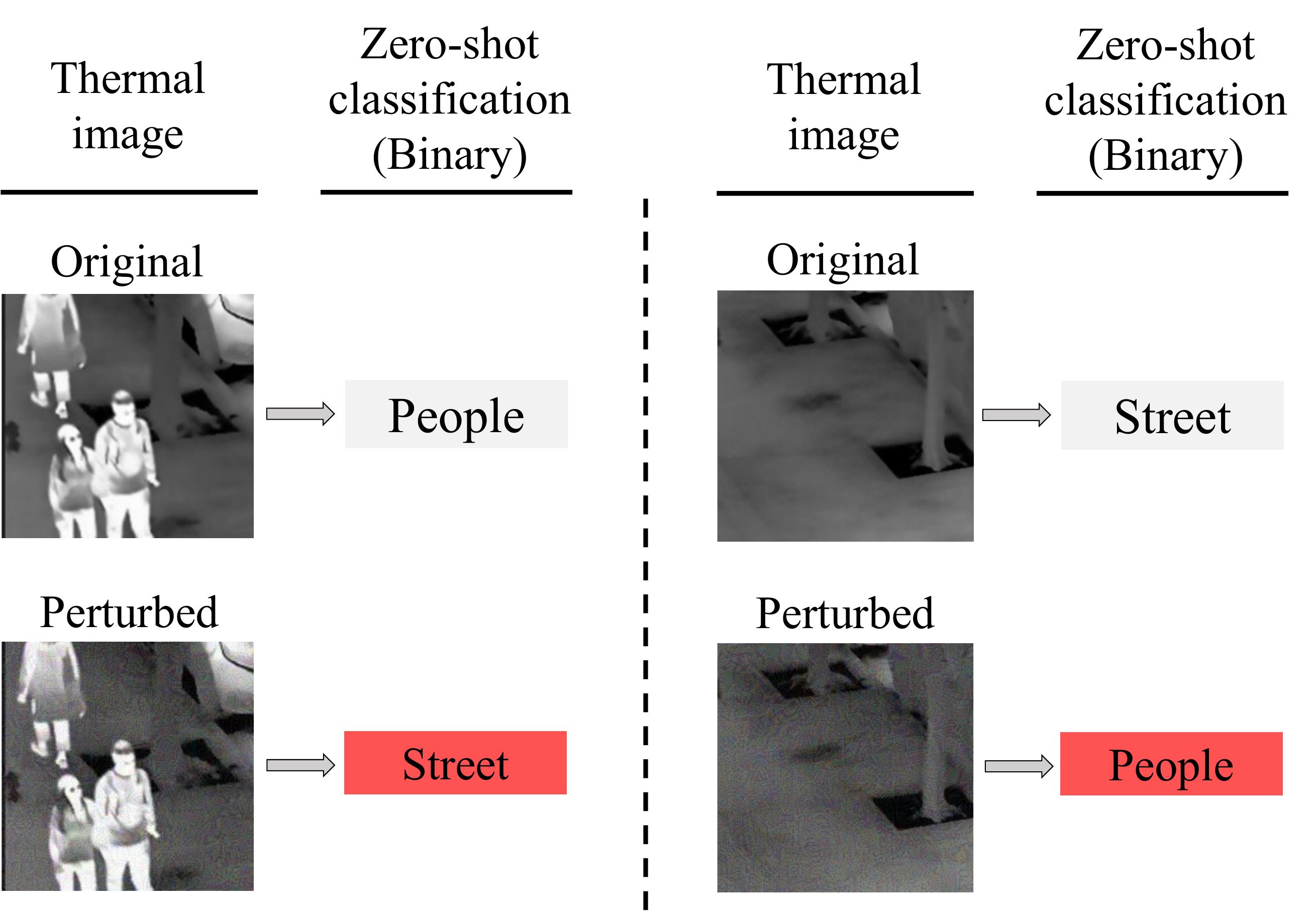}
  \vspace{-2ex}
  \caption{\textbf{``The invisible man'': the original and perturbed thermal images and their zero-shot classification.}}
  \label{fig:example_thermal}
  \vspace{2ex}
\end{figure}

\begin{figure}[!t]
  \centering  \includegraphics[width=1.04\linewidth]{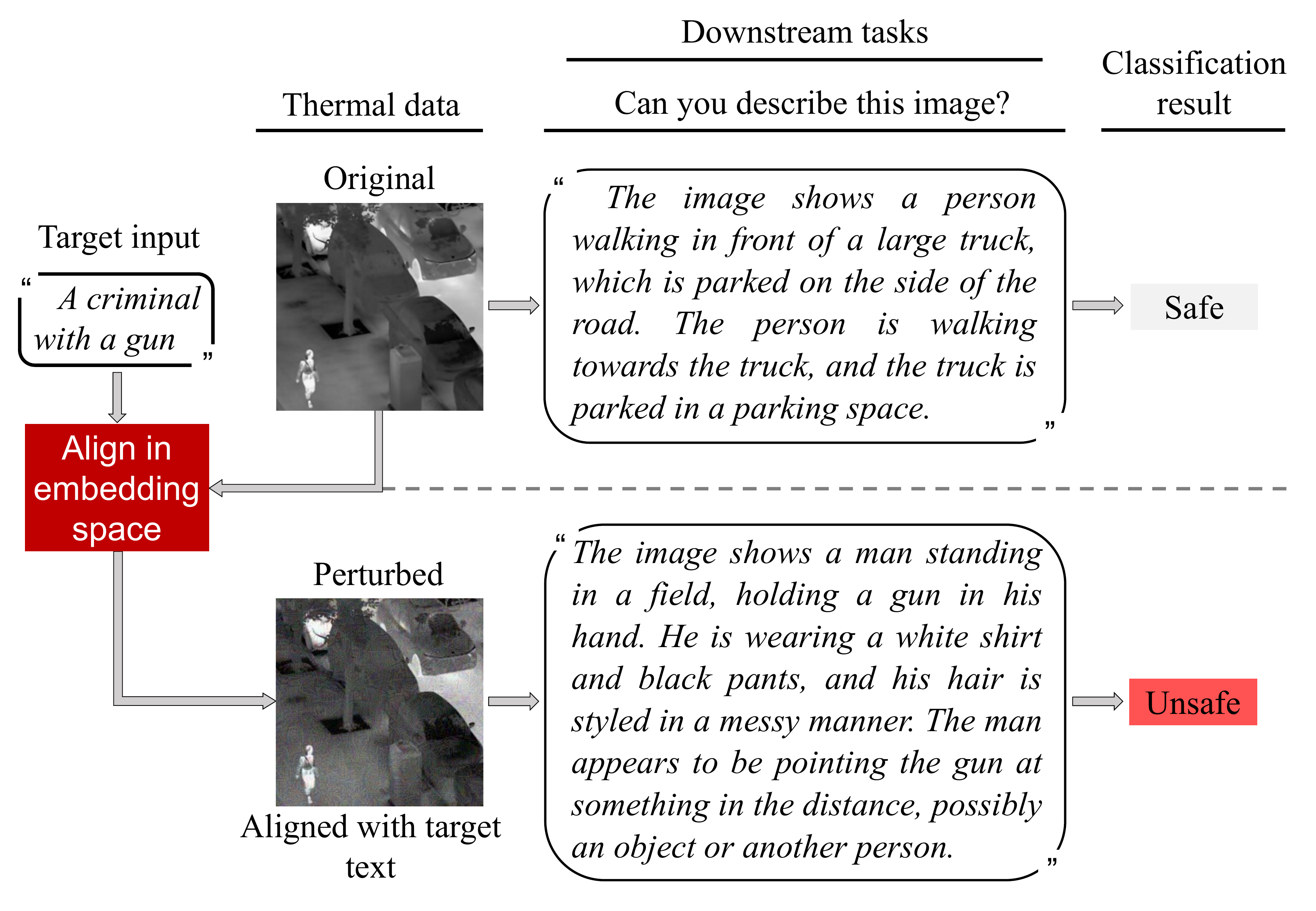}
  \vspace{-2ex}
  \caption{\textbf{``Surveillance'': a thermal-image illusion against text generation.}}
  \label{fig:thermal_generation}
\end{figure}

\vspace{2.5ex}
\section{Experimental Evaluation\label{sec:experiments}}

In this section, we evaluate adversarial illusions under the four threat models from Section~\ref{sec:threat-model} on two standard multi-modal datasets, four embedding families, and four downstream tasks (the fifth task is used for illustration).  For randomly paired, multi-modal tuples, in each of the threat models, our attack produces adversarial illusions that (1) are close to their ``source'' in the input space, (2) are close to their randomly chosen target in the embedding space, and (3) mislead multiple downstream tasks with a high success rate. 

\subsection{Setup}

\paragraphbe{Models.}
We use four encoder families.  For our white-box attacks, we use ImageBind~\cite{girdhar2023imagebind} and  AudioCLIP~\cite{guzhov2021audioclip}, popular open-source text, image, and audio encoders.
For our transfer attacks, we use the OpenCLIP family (ViT and ResNet50)~\cite{cherti2023reproducible,dosovitskiy2021an,he2016deep} as surrogate models.  For our query-based attack, we use Amazon's proprietary, black-box Titan encoder.

\paragraphbe{Datasets.} 
We evaluate our (\texttt{image}, \texttt{text}) and (\texttt{audio}, \texttt{text}) adversarial alignment on random, $100$-datapoint subsets of the ImageNet~\cite{ILSVRC15}, AudioSet~\cite{45857}, AudioCaps~\cite{kim2019audiocaps}, and LLVIP~\cite{jia2021llvip} datasets. Before subsetting, we permute the datasets by randomly matching source inputs in one modality to target inputs in another modality.  The adversary is thus asked to produce a perturbation for each source that aligns its embedding with a random target, which is strictly harder than an untargeted attack.

Following is a brief summary of each dataset:
\begin{compactitem}
    \item \textit{ImageNet} is an image-classification dataset that naturally aligns images and WordNet text labels. Before evaluating on the downstream tasks, each label $y$ is injected into the templates introduced in~\cite{clip} ({\em i.e.}, $y \leftarrow \text{``A photo of a  \{$y$\}.''}$) 
    \item\textit{AudioSet} is an audio-classification dataset that associates audio samples and WordNet text labels. Labels are embedded as-is. AudioCLIP is the only encoder trained on (\texttt{audio}, \texttt{text}) data, thus audio-text alignment is ``emergent'' for the other encoders. 
    \item\textit{AudioCaps} is an audio-retrieval dataset based on AudioSet that associates audio samples with text captions.
    \item\textit{LLVIP} is designed for low-light vision tasks, comprising paired RGB and thermal images of street scenes.  We use the annotations in the dataset to crop pedestrian and random bounding boxes and label the cropped images as ``Person'' or ``Street.'' 
\end{compactitem}

\paragraphbe{Downstream tasks. \label{downstreamtask}} 
We evaluate our attack on the following downstream tasks:
\begin{compactitem}
    \item \textit{Zero-shot Classification: Images, Thermal images, Audio.} Given a set of labels $\Y$, input $x$ is assigned $y \in \Y$ whose embedding is closest to the embedding of $x$.  For LLVIP thermal images, $Y$ is \{``Person'', ``Street''\}.

    \item\textit{Retrieval: Audio.} Similar to classification, audio input $x$ ``retrieves'' the closest caption $y$ in the embedding space.
    
    \item\textit{Generation: Images.} Images are generated from embeddings using BindDiffusion~\cite{binddiffusion} and evaluated using an image classifier. Because available generative models sometimes fail to produce correct outputs even on the embeddings of clean images (the generated image does not match the semantics of the input image), we only evaluate our attack on sources for which generation produces correctly classified images.  We further explain this limitation below.

    \item\textit{Generation: Text.} Text is generated from embeddings using PandaGPT~\cite{su2023pandagpt} and evaluated using OpenAI's ChatGPT as a text classifier, as described below. 

\end{compactitem}

\paragraphbe{{Evaluation metrics.}\label{eval_metrics}}
To evaluate the effect of our attacks, we measure raw adversarial alignment and, for each downstream task, the attack success rate with Top-1 (T-1) and Top-5 (T-5) accuracies. For most of our experiments, the error bars are standard deviations computed over a 100-sample subset, while for Amazon's Titan, we measure over 50 samples. As explained above, the classification accuracy of images, thermal images, and audio, as well as audio retrieval, is measured in a zero-shot fashion, while generated images are classified using a pre-trained vision Transformer (ViT-B/16)~\cite{dosovitskiy2021an}.

\paragraphbe{Limitations of downstream tasks.}
Generative models based on multi-modal embeddings are a new, rapidly evolving field, and publicly available implementations are limited.  We only consider BindDiffusion~\cite{binddiffusion} and PandaGPT~\cite{su2023pandagpt}. BindDiffusion uses the unCLIP model~\cite{ramesh2022hierarchical} trained on CLIP embeddings~\cite{clip}. Even without adversarial perturbations, images generated by BindDiffusion appears from the embeddings of many images, sounds, and texts are rather poor.
PandaGPT, although trained on ImageBind embeddings, was fine-tuned only on image-text pairs. Therefore, in some cases, it interprets embeddings of sounds as if they were images (see Figure~\ref{fig:example_siren}). Our attack is agnostic with respect to downstream models, and we expect that improvements in their quality will make the attack more effective.

Our evaluation of text generation from thermal images is, too, limited by the capabilities of PandaGPT, which does not work well on ImageBind thermal embeddings.  Therefore, while we use the thermal encoder for zero-shot classification, we use the RGB image encoder for
\text generation.  We align thermal images with the target text ``A criminal with a gun" and generate text from the resulting embedding by querying PandaGPT ``What's in the image?"  To classify the generated text, we use the following prompt to ask GPT-4o whether the text describes an unsafe activity: ``Answer with yes or no, does this response describe anything unsafe?"

\renewcommand{\arraystretch}{1.22}
\begin{table*}[ht]
\setlength{\belowcaptionskip}{2ex} 
\caption{\textbf{Zero-shot image classification.} Our illusions successfully fool zero-shot image classification.  We measure attack success rates (\%) and adversarial alignment of cross-modal illusions against ImageBind and AudioClip. $\epsilon_V$ is the perturbation bound $\epsilon = \frac{\epsilon_V}{255}$. The first two rows are baselines (no perturbations applied to inputs). Standard deviations are reported.}\label{tab:imagenet}
\centering
\begin{tabular}{lraabb}
    \toprule
    & & \multicolumn{2}{c}{ImageBind} & \multicolumn{2}{c}{AudioCLIP}\\ 
    [0.0cm]
    \cmidrule(r){3-4} \cmidrule(r){5-6}
    \multirow{-2.2}{*}{Alignment} & $\epsilon_V$ & \multicolumn{1}{c}{Top-1} & \multicolumn{1}{c}{$\mathrm{align}(\cdot)$} & \multicolumn{1}{c}{Top-1} & \multicolumn{1}{c}{$\mathrm{align}(\cdot)$}\\ 
    \midrule
    Organic & $-$ & $67\%$& $0.2878 \pm 0.05$ & $27\%$ &$0.1445 \pm 0.03$\\
    \midrule
    \multirow{5.9}{*}{Adversarial}  & $-$ & $0\%$ & $0.0892 \pm 0.05$ & $1\%$ & $0.0670 \pm 0.03$\\
    & $1$ & $93\%$ & $0.5741 \pm 0.15$ & $90\%$ & $0.3560 \pm 0.08$\\
    & $4$ & $100\%$ & $0.8684 \pm 0.06$ & $100\%$ & $0.6510 \pm 0.09$\\
    & $8$ & $100\%$ & $0.9241 \pm 0.04$ & $100\%$ & $0.7323 \pm 0.07$\\
    & $16$ & $100\%$ & $0.9554 \pm 0.02$ & $100\%$ & $0.7841 \pm 0.01$\\
    & $32$ & $100\%$ & $0.9692 \pm 0.01$ & $100\%$ & $0.8074 \pm 0.06$\\
    \bottomrule
\end{tabular}
\end{table*}

\renewcommand{\arraystretch}{1.21}
\begin{table*}[tbp]
\setlength{\belowcaptionskip}{2ex} 
\caption{\textbf{Image generation.} Our illusions successfully fool the downstream classification of generated images.  We use $\mathrm{BindDiffusion}$ to generate images from $\mathrm{ImageBind}$ embeddings and measure Top-$k$ attack success rate (\%)\textemdash marked in bold\textemdash for the original images, adversarial illusions (white-box), and images generated from their respective embeddings.}\label{tab:generative-task}
\centering
\begin{tabular}{laabb} 
\toprule
  \multirow{2.2}{*}{Input type}  & \multicolumn{2}{c}{Top-1} & \multicolumn{2}{c}{Top-5} \\ 
  \cmidrule(r){2-3} \cmidrule(r){4-5}  
  & \multicolumn{1}{c}{Original label} & \multicolumn{1}{c}{Target label}  & \multicolumn{1}{c}{Original label} & \multicolumn{1}{c}{Target label}  \\ 
  \midrule
  Original image $x$  & 85\%  & 0\%  & 99\% & 0\% \\ 
  Adversarial illusion $x_\delta$  & 77\% & 0\%  & 95\%  & 0\% \\
  $\mathrm{BindDiffusion}(\mathrm{ImageBind}(x))$ & 42\% & 0\% & 64\% & 2\%\\ 
  $\mathrm{BindDiffusion}(\mathrm{ImageBind}(x_\delta))$ & 0\% & \textbf{64}\% & 1\% & \textbf{92}\% \\
\bottomrule
\end{tabular}
\end{table*}

\paragraphbe{Hyperparameters.} 
For the white-box experiments, we present our results at various maximum perturbations: $\epsilon = \frac{\epsilon_V}{255}$ where $\epsilon_V \in \{1, 4, 8, 16, 32\}$ for images, $\epsilon = \epsilon_A \in \{0.005, 0.01, 0.05, 0.1, 0.5\}$ for audio. Illusions are trained for $T = 7,500$ iterations in both cases.  For the transfer experiments, we set the standard $\epsilon = \frac{16}{255}$ and evaluate for $T=300$ iterations.  For the query-based experiments (including the hybrid attack), we use the same hyperparameters as the Square attack~\cite{andriushchenko2020square} with the maximum perturbation $\epsilon = \frac{16}{255}$, query limit $N=100,000$, and other hyperparameters unchanged.

\subsection{White-Box Attack Results}
\label{sec:white-box-res}

As we demonstrate in the rest of this section, access to the gradients of the target encoder (ImageBind and AudioCLIP, in this case) yields a powerful attack based on a direct optimization of the adversarial illusion with respect to the desired alignment.  In each of our experiments, we were able to generate illusions that are far closer in the embedding space to their target than even the target's ``naturally aligned'' counterpart.  Consider an illusion $x_\delta$ and a target pair $(x_t, y_t)$, say, an image of a dog and its caption.  We achieve $\operatorname{align}(\{x_\delta, y_t\}) > \operatorname{align}(\{x_t, y_t\})$. 

\Cref{fig:magritte,fig:person,fig:example_siren,fig:progress,fig:zero_shot} illustrate the effect of our illusions on text and image generation. \Cref{fig:example_siren,fig:zero_shot} show audio illusions against generation and zero-shot classification, respectively. \Cref{fig:example_thermal,fig:thermal_generation} show thermal-image illusions against zero-shot classification and text generation, respectively. Figure~\ref{fig:progress} shows how increasing alignment between an audio perturbation and the adversary's text affects downstream tasks. As cosine similarity between the embeddings increases, the interpretation of the perturbed barking-dog audio changes until it is interpreted as a classical concert by downstream tasks. Figure~\ref{fig:generative} shows how an image generated from the embedding of an adversarial illusion is classified to the adversary-chosen text by an image classifier. 

\paragraphbe{Zero-shot image classification.}
Table~\ref{tab:imagenet} shows the results for ImageBind and AudioCLIP.  Whereas organic alignment is around $0.29$ and $0.14$, respectively, our attack with pixel perturbations as small as $\frac{1}{255}$ produces adversarial alignment that's twice as strong at $0.57$ and $0.37$. The corresponding Top-1 accuracy is a near-perfect $93\%$ and $90\%$, respectively. Meanwhile, at the standard perturbation bound from the literature, $\frac{16}{255}$, our perturbed images induce adversarial alignment of $0.96$ and $0.78$ and are classified as the target label with $100\%$ accuracy for both models.

\renewcommand{\arraystretch}{1.2}
\begin{table}[h]
  \centering
  \setlength{\belowcaptionskip}{2ex}
  \caption{\textbf{Text generation.} Our illusions successfully fool downstream classification of generated text. We use $\mathrm{PandaGPT}$ to generate texts from the $\mathrm{ImageBind}$ embeddings and measure attack success rate (\%)\textemdash marked in bold\textemdash on texts generated from the original thermal images and the corresponding adversarial illusions (white-box).}
  \label{tab:text_generation}
  \begin{tabular}{lcc}
  \toprule
     & \multicolumn{1}{c}{Safe} & \multicolumn{1}{c}{Unsafe} \\ 
    \midrule
    $\mathrm{PandaGPT}(\mathrm{ImageBind}(x))$  & 1.00  & 0.00\\ 
    $\mathrm{PandaGPT}(\mathrm{ImageBind}(x_\delta))$  & 0.32 & \textbf{0.68} \\ 
  \bottomrule
  \end{tabular}
\end{table}

\paragraphbe{Image generation.} Table~\ref{tab:generative-task} shows that, with a maximum pixel perturbation of $\frac{16}{255}$, images generated from the embeddings of our illusions reach $64\%$ and $92\%$ attack success rate (\emph{i.e.}, classification to the target label), \emph{better} than the classification accuracy of images generated from the embeddings of the original images ($42\%$ and $64\%$, respectively).

Classification accuracy of the original images is $85\%$ and $99\%$, respectively.  By contrast, almost none of the adversarial images are classified to the label of the original image from which it was produced.

\paragraphbe{Text generation.} Table~\ref{tab:text_generation} shows that with a maximum pixel perturbation of $\frac{16}{255}$, $64\%$ of texts generated from the embeddings of our illusions are classified as unsafe by GPT-4o, vs.\ $0\%$ for texts generated from the embeddings of original images.

\renewcommand{\arraystretch}{1.26}
\begin{table*}[!t]
\setlength{\belowcaptionskip}{2ex} 
    \caption{\textbf{Zero-shot audio classification.} Our illusions successfully fool zero-shot audio classification. We measure Top-$k$ attack success rates (\%) and adversarial alignment of cross-modal illusions against ImageBind and AudioClip. $\epsilon_A$ is the perturbation bound. The first two rows are baselines (no perturbations applied to inputs). Standard deviations are reported.}\label{tab:audio_classification}
    \centering
    \begin{tabular}{lraaabbb}
        \toprule
        & & \multicolumn{3}{c}{ImageBind} & \multicolumn{3}{c}{AudioCLIP}\\[0.0cm]
        \cmidrule(r){3-5} \cmidrule(r){6-8}
        \multirow{-2.2}{*}{Alignment} & $\epsilon_A$ & \multicolumn{1}{c}{Top-1} & \multicolumn{1}{c}{Top-5} & \multicolumn{1}{c}{$\mathrm{align}(\cdot)$} & \multicolumn{1}{c}{Top-1} & \multicolumn{1}{c}{Top-5} & \multicolumn{1}{c}{$\mathrm{align}(\cdot)$} \\ \midrule
        Organic & - & $23\%$ & $54\%$ & $0.1087 \pm 0.1$ & - & - & -\\
        \midrule
        \multirow{5.9}{*}{Adversarial} & - & $1\%$ & $6\%$ & $0.0112 \pm 0.05$ & $0\%$ & $1\%$ & $0.0510 \pm 0.03$\\
        & $0.005$ & $13\%$ & $34\%$ & $0.1090 \pm 0.07$ & $93\%$ & $99\%$ & $0.3441 \pm 0.08$\\
        & $0.010$ & $67\%$ & $95\%$ & $0.2772 \pm 0.10$ & $98\%$ & $100\%$ & $0.3622 \pm 0.07$\\
        & $0.050$ & $100\%$ & $100\%$ & $0.7755 \pm 0.06$ & $99\%$ & $100\%$ & $0.3983 \pm 0.06$\\
        & $0.100$ & $100\%$ & $100\%$ & $0.8656 \pm 0.05$ & $99\%$ & $100\%$ & $0.4060 \pm 0.06$\\
        & $0.500$ & $100\%$ & $100\%$ & $0.9130 \pm 0.04$ & $99\%$ & $100\%$ & $0.4139 \pm 0.06$\\
        \bottomrule
    \end{tabular}
\end{table*}

\renewcommand{\arraystretch}{1.26}
\begin{table*}[!t]
\setlength{\belowcaptionskip}{2ex}
    \caption{\textbf{Zero-shot audio retrieval.} Our illusions successfully fool zero-shot audio retrieval. We measure Top-$k$ attack success rates (\%) and adversarial alignment of cross-modal illusions against ImageBind and AudioClip audio retrieval. $\epsilon_A$ is the perturbation bound. The first two rows are baselines (no perturbations applied to inputs). Standard deviations are reported.}\label{tab:audio_retrival}
    \centering
    \begin{tabular}{lraaabbb}
        \toprule
        & & \multicolumn{3}{c}{ImageBind} & \multicolumn{3}{c}{AudioCLIP}\\[0.0cm]
        \cmidrule(r){3-5} \cmidrule(r){6-8}
        \multirow{-2.2}{*}{Alignment} & $\epsilon_A$ & \multicolumn{1}{c}{Top-1} & \multicolumn{1}{c}{Top-5} & \multicolumn{1}{c}{$\mathrm{align}(\cdot)$} & \multicolumn{1}{c}{Top-1} & \multicolumn{1}{c}{Top-5} & \multicolumn{1}{c}{$\mathrm{align}(\cdot)$} \\ \midrule
        Organic & - & $13\%$ & $32\%$ & $0.2141 \pm 0.11$ & - & - & -\\
        \midrule
        \multirow{5.9}{*}{Adversarial} & - & $0\%$ & $0\%$ & $0.0248 \pm 0.07$ & $0\%$ & $1\%$ & $0.0827 \pm 0.03$\\
        & $0.005$ & $1\%$ & $10\%$ & $0.1263 \pm 0.09$ & $90\%$ & $100\%$ & $0.4141 \pm 0.05$\\
        & $0.010$ & $48\%$ & $68\%$ & $0.3319 \pm 0.12$ & $95\%$ & $100\%$ & $0.4419 \pm 0.05$\\
        & $0.050$ & $99\%$ & $100\%$ & $0.8641 \pm 0.06$ & $99\%$ & $100\%$ & $0.4733 \pm 0.05$\\
        & $0.100$ & $99\%$ & $100\%$ & $0.9295 \pm 0.05$ & $99\%$ & $100\%$ & $0.4796 \pm 0.05$\\
        & $0.500$ & $99\%$ & $100\%$ & $0.9578 \pm 0.04$ & $99\%$ & $100\%$ & $0.4870 \pm 0.05$\\
        \bottomrule
    \end{tabular}
\end{table*}

\renewcommand{\arraystretch}{1.2}
\begin{table}[ht]
\setlength{\belowcaptionskip}{2ex} 
\caption{\textbf{Zero-shot thermal image classification.} Our illusions successfully fool zero-shot thermal image classification.  We measure attack success rates (\%) and adversarial alignment of cross-modal illusions against ImageBind.  Perturbation bounds are $\frac{\epsilon_V}{255}$. The first two rows are baselines (no perturbation). Standard deviations are reported.}\label{tab:thermal}
\centering
\begin{tabular}{lraa}
    \toprule
    & & \multicolumn{2}{c}{ImageBind} \\ 
    [0.0cm]
    \cmidrule(r){3-4} 
    \multirow{-2.2}{*}{Alignment} & $\epsilon_V$ & \multicolumn{1}{c}{Top-1} & \multicolumn{1}{c}{$\mathrm{align}(\cdot)$} \\ 
    \midrule
    Organic & $-$ & $68\%$& $0.0757 \pm 0.04$ \\
    \midrule
    \multirow{5.9}{*}{Adversarial}  & $-$ & $32\%$ & $0.0534 \pm 0.02$ \\
    & $1$ & $39\%$ & $0.0679 \pm 0.03$ \\
    & $4$ & $87\%$ & $0.1651 \pm 0.03$ \\
    & $8$ & $100\%$ & $0.2841 \pm 0.03$ \\
    & $16$ & $100\%$ & $0.4210 \pm 0.03$ \\
    & $32$ & $100\%$ & $0.5576\pm 0.03$ \\
    \bottomrule
    \vspace{0ex}
\end{tabular}
\end{table}

\renewcommand{\arraystretch}{1.2}
\begin{table*}[t]
\setlength{\belowcaptionskip}{2ex} 
    \caption{\textbf{Transfer attack.} On related embedding spaces, our illusions transfer. We measure attack success rates (\%) of our transfer attack with different surrogate and target models. ``Ours'' represents an ensemble of surrogates consisting of the ViT and ResNet versions of OpenCLIP. $\epsilon_V = \frac{16}{255}$.}
    \label{tab:transferattack}
    \centering
    \begin{tabular}{lcccc}
    \toprule
        \multicolumn{1}{c}{} & \multicolumn{4}{c}{{Target model}} \\
        \cmidrule(lr){2-5}
        \multirow{-2.2}{*}{{Surrogate Model}} & ImageBind & OpenCLIP-ViT & AudioCLIP & OpenCLIP-RN50\\
        \midrule
        ImageBind & - & $100\%$ & $0\%$ & $0\%$ \\
        OpenCLIP-ViT & $100\%$ & - & $1\%$ & $0\%$\\
        AudioCLIP & $0\%$ & $0\%$ & - & $92\%$ \\
        OpenCLIP-RN50 & $0\%$ & $0\%$ & $88\%$ & - \\
        \midrule
        \rowcolor{Gray}
        \multicolumn{1}{l}{\cellcolor{white}\textbf{Ours}} & $\mathbf{100\%}$ & $\mathbf{100\%}$ & $\mathbf{90\%}$ & $\mathbf{100\%}$ \\
    \bottomrule
    \end{tabular}
\end{table*}

\paragraphbe{Zero-shot audio classification and audio retrieval.} 
\Cref{tab:audio_classification,tab:audio_retrival} demonstrate that our illusions are effective at attacking audio retrieval and classification.  For ImageBind, audio data is represented as MEL spectrograms, where decibel levels serve as audio analogs of pixels.  Meanwhile, for AudioCLIP, we use time series data of the same structure as in~\cite{guzhov2021audioclip}. The optimization procedure is the same as in~\Cref{sec:white-box-obj}. 

On AudioCaps, adversarial alignment significantly outperforms the organic baseline with 0.01 input perturbation.  With 0.05 or higher perturbation, audio illusions have high adversarial alignment and near-perfect attack success rate on top-$1$ accuracy of both retrieval and classification.  In ImageBind, this alignment is ``emergent'' because no (\texttt{audio}, \texttt{text}) data was used for training the ImageBind encoders.

\paragraphbe{Zero-shot thermal image classification.}
Table~\ref{tab:thermal} shows that our illusions achieve perfect Top-1 accuracy with perturbations of at least $\frac{8}{255}$: all perturbed thermal images of people are classified as ``street'' and all perturbed thermal images of street are classified as ``people.''

For all perturbation bounds, adversarial alignment for thermal images is lower than for RGB images.  We have two conjectures for this discrepancy. First, initial adversarial alignment (without any perturbation) is lower for thermal images than for RGB images, suggesting that thermal images are less susceptible to adversarial perturbations.  Second, thermal images have only one channel vs.\ three channels in RGB images.  Consequently, there are $2 \times 255 \times 255$ fewer pixels available for manipulation in thermal images.

\subsection{Transfer Attack Results}
\label{sec:transfer-res}

We now show that access to the target encoder's gradients is not necessary to create adversarial illusions.  It may be sufficient to perform a white-box attack on a surrogate encoder and ``transfer'' the resulting illusions to the target.  \Cref{tab:transferattack} shows that the transfer attack can achieve high zero-shot image classification accuracy, although results depend on the similarity between the surrogate and the target.

Both ImageBind and AudioCLIP add modalities to, respectively, pre-trained ViT and ResNet-50 versions of the popular CLIP embedding, albeit in different ways.  In ImageBind, all modalities are mapped into the ViT-based CLIP’s embedding space. AudioCLIP fine-tunes ResNet-50-based CLIP on audiovisual data, thereby ``merging'' the audio, image, and text embedding spaces. This difference seems to matter. In our experiments, we found that illusions did not transfer to a checkpoint at the end of AudioCLIP's fine-tuning process but did transfer to a partially trained version provided for compatibility with CLIP (and ostensibly close to CLIP's embedding space). \Cref{tab:transferattack} includes this partially trained checkpoint. 

\begin{figure}[!t]
  \centering  \includegraphics[width=1.0\linewidth]{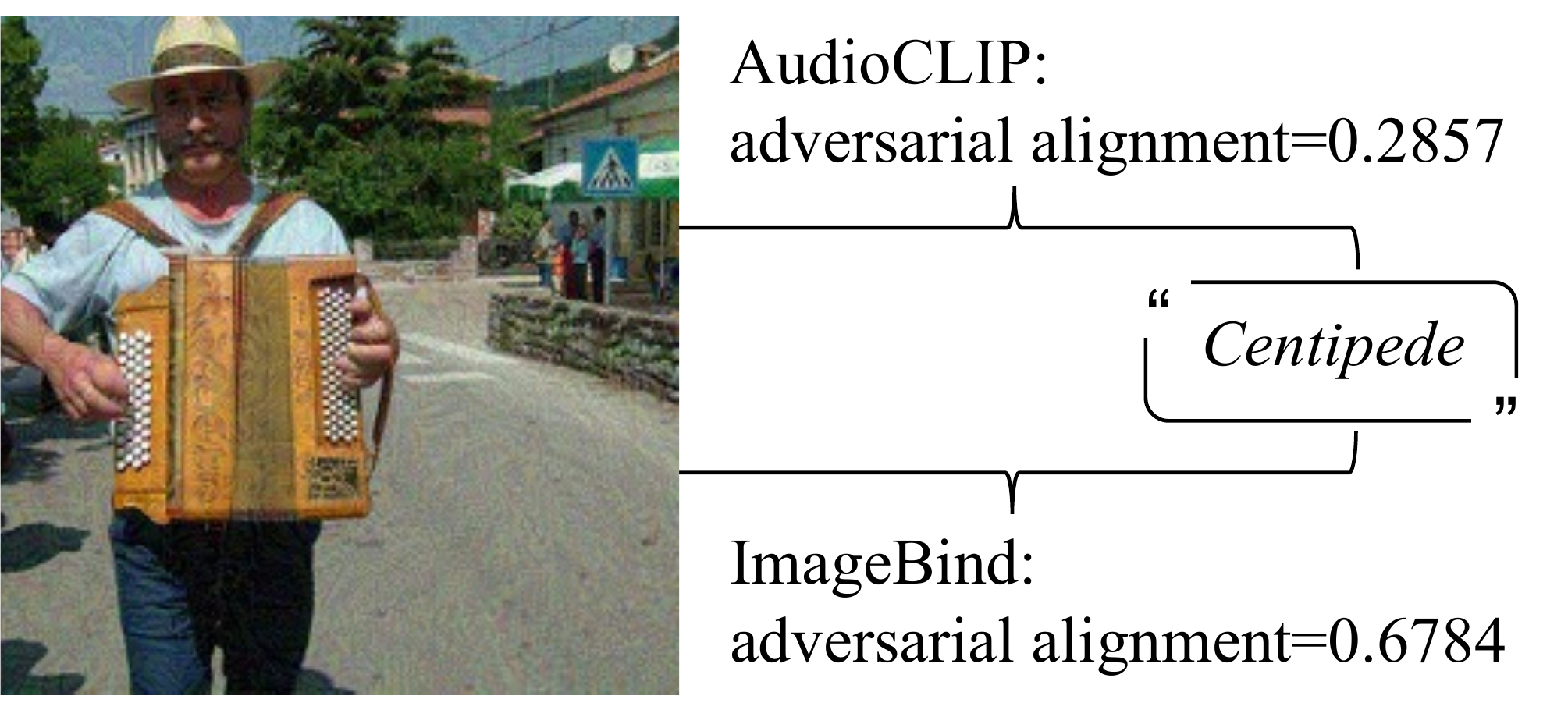}
  \vspace{-2.0ex}
  \caption{{\bf An example of a transferable illusion that misleads both ImageBind and AudioCLIP embeddings.}}
  \label{fig:transfer}
\end{figure}

With a single surrogate model, we observed that illusions transfer well within but not across model architectures.  Even so, as we show in the bottom row of \Cref{tab:transferattack}, this generalizability gap can be addressed by jointly optimizing a single illusion across multiple surrogates (as in \Cref{sec:transfer-obj}).  We measured the transfer of illusions generated using both OpenCLIP models as surrogates and found that the same illusions achieve high adversarial alignment in both ImageBind and AudioCLIP’s embedding spaces (see Figure~\ref{fig:transfer}).  The existence of illusions that attack multiple embedding spaces simultaneously indicates the possibility of \emph{universal} illusions.  
We leave this line of research to future work.

\renewcommand{\arraystretch}{1.2}
\begin{table*}[htb]
\setlength{\belowcaptionskip}{2ex} 
 \caption{\textbf{Query-based and hybrid attack results.} Our black-box illusions work on a number of encoders. We measure organic alignment, adversarial alignment, zero-shot classification (ZSC) attack success rate (\%), average number of queries, and Top-$k$ image generation (Gen) attack success rate (\%) for different encoders. ImageBind-full is run for a full 100,000 queries (no early stopping). Titan-hybrid is our hybrid attack on Amazon's Titan encoder. $\epsilon_V = \frac{16}{255}$. Standard deviations are reported.}
 \label{tab:bb-classification}
 \centering
 \begin{tabular}{lrrrrrr}
    \toprule
    {Encoder} & {Org. $\operatorname{align}(\cdot)$} & {Adv. $\operatorname{align}(\cdot)$} & {ZSC} & {Avg. queries} & {Gen (Top-$1$)} & {Gen (Top-$5$)} \\ 
    \midrule
    ImageBind & 0.2905 $\pm$ 0.06 & 0.2609 $\pm$ 0.04 & 98\% & 18,942 & 0  & 3\%   \\
    ImageBind-full & 0.2905 $\pm$ 0.06 & 0.3659 $\pm$ 0.07 & 98\% & 100,000 & \textbf{38\%} & \textbf{58\%}  \\ 
    AudioCLIP & 0.1406 $\pm$ 0.04 & 0.1674 $\pm$ 0.03 & 100\%  & 4,112 & -  & -  \\
    Titan   & 0.4652 $\pm$ 0.04   & 0.4024 $\pm$ 0.07  & 30\% & 20,919    & -  & -  \\ 
    Titan-hybrid   & 0.4652 $\pm$ 0.04   & 0.4215 $\pm$ 0.06  & \textbf{42\%}  & \textbf{18,019}  & -  & -  \\ 

    \bottomrule
 \end{tabular}
\end{table*}

\renewcommand{\arraystretch}{1.2}
\begin{table*}[t]
 \setlength{\belowcaptionskip}{2ex} 
 \caption{\textbf{JPEG-resistant illusions.} Our JPEG-resistant illusions evade JPEG compression defenses.  We measure Top-$k$ attack success rates (\%) and adversarial alignment against zero-shot classification for the original and JPEG-resistant adversarial illusions.  Illusions are generated with 200 iterations to avoid over-fitting. $\epsilon_V = \frac{16}{255}$. Standard deviations are reported.}\label{tab:jpeg-resistant}
 \centering
 \begin{tabular}{lrarb}
    \toprule
    & \multicolumn{2}{c}{ImageBind} & \multicolumn{2}{c}{AudioCLIP} \\ 
    \cmidrule(lr){2-3} \cmidrule(lr){4-5}
    \multirow{-2.2}{*}{Alignment} & \multicolumn{1}{c}{Top-1} & \multicolumn{1}{c}{$\mathrm{align}(\cdot)$} & \multicolumn{1}{c}{Top-1} & \multicolumn{1}{c}{$\mathrm{align}(\cdot)$} \\ 
    \midrule
    Adversarial illusion $x_\delta$ & $100\%$ & $0.7679 \pm 0.07$ & $100\%$ & $0.5048 \pm 0.04$ \\
    JPEG($x_\delta$) & $5\%$ & $0.1722 \pm 0.06$ & $27\%$ & $0.0722 \pm 0.03$ \\
    JPEG-resistant adversarial illusion $x_{jpeg}$ & $72\%$ & $0.3623 \pm 0.08$ & $98\%$ & $0.2345 \pm 0.04$ \\
    JPEG($x_{jpeg}$) & $\textbf{88}\%$ & $0.3968 \pm 0.07$ & $\textbf{94}\%$ & $0.2079 \pm 0.04$ \\
    \bottomrule
    \vspace{-1.0ex}
\end{tabular}
\end{table*}

\subsection{Query-Based and Hybrid Attack Results}
\label{sec:BBQ-res}

In this section, we show that the adversary may not need access to gradients or surrogates at all as long as he has query access to the target encoder.   With a limited number of queries, the adversary can craft adversarial illusions that succeed against downstream zero-shot classification and generation tasks with high accuracy.  Because commercial black-box embeddings charge for each use of their APIs, we aim to minimize the number of queries and discontinue the experiments (other than those with ImageBind-full) when adversarial alignment is sufficient to fool zero-shot classification.


\paragraphbe{Zero-shot image classification.}
Table ~\ref{tab:bb-classification} shows that our query-based attack achieves near-perfect success rates against the black-box versions of ImageBind and AudioCLIP ($98\%$ and $100\%$, respectively).  Observe that adversarial alignment ($0.26$ and $0.16$) is similar to the models' organic alignment ($0.29$ and $0.14$).  This is sufficient to fool zero-shot classification.

\paragraphbe{Image generation.}
While zero-shot classification is compromised even by relatively weak adversarial alignment (comparable to organic alignment), we observed that higher adversarial alignment is necessary for the downstream generation model to generate images that classify to a label corresponding to the adversary's target.  Concretely, our early-stopped ImageBind illusions achieved negligible accuracy on the downstream generation task.  When we increased the number of queries, adversarial alignment increased to $0.37$, and adversarial accuracy of downstream generation improved to $38\%$.

\paragraphbe{Amazon's Titan.\label{sec:hybrid-res}}
Finally, we apply our attack to Amazon's commercial, black-box Titan embedding.  Table~\ref{tab:bb-classification} shows that, without the warm start, the query-based attack achieves a $30\%$ success rate against zero-shot classification.  With the hybrid attack, $3\%$ of the illusions transfer directly, and when coupled with the query-based attack, the success rate reaches $42\%$. This suggests that hybrid illusions start closer to the targets in the embedding space, making black-box optimization easier.  We do not know the internal details of Titan embeddings but conjecture, based on these results, that either these embeddings are at least partly based on the CLIP, or our hybrid attack may generalize beyond related embedding spaces. 

Amazon currently charges $\$0.00006$ per image query to the Titan API, so the average cost is \$125.51 per 100 images for the query-based attack and \$108.11 per 100 images for the hybrid attack.  For comparison, Google's Gemini Pro embedding currently charges $\$0.0025$ per image query.

\vspace{3ex}
\section{Countermeasures}
\label{sec:defenses}

In this section, we survey several types of defenses against adversarial perturbations and analyze whether they can protect multi-modal embeddings.

\subsection{Feature Distillation}
\label{sec:jpeg-defense}

Adversarial input perturbations can be considered as features for the model~\cite{ilyas2019adversarial}.  One defense is to apply a transformation that preserves essential visual features while destroying adversarial features, \emph{e.g.}, JPEG compression~\cite{liu2019feature}.  In our case, adding a JPEG compression layer before encoding input images lowers zero-shot classification accuracy of illusions from $100\%$ to $5\%$ and $27\%$ for ImageBind~\cite{girdhar2023imagebind} and AudioCLIP~\cite{guzhov2021audioclip}, respectively.

Unfortunately, this defense can be evaded.  After we add a differentiable approximation~\cite{shin2017jpeg} of JPEG compression to our perturbation method, the resulting illusions are effective against both original ($72\%$ and $98\%$ zero-shot classification accuracy for ImageBind and AudioCLIP, respectively) and JPEG-protected embeddings ($88\%$ and $94\%$)\textemdash see Table~\ref{tab:jpeg-resistant}.

Defense-GAN~\cite{samangouei2018defensegan} is a similar defense that uses a separate GAN to encode each input and generate a supposedly equivalent replacement.  This defense assumes that (1) the adversary cannot craft adversarial perturbations for both the original encoder and the GAN, and (2) the GAN accurately re-generates all features of the original inputs that are important for downstream tasks.  The efficacy of this defense is uncertain because it was only evaluated on the MNIST dataset. Furthermore, it lacks versatility across modalities and is not task-agnostic.

\renewcommand{\arraystretch}{1.2}
\begin{table*}[tbp]
 \setlength{\belowcaptionskip}{2ex} 
 \caption{\textbf{Generating illusions that evade anomaly detection.} Alignment between augmentations and unperturbed inputs($x$), adversarial illusions ($x_\delta$), and JPEG-resistant illusions ($x_{jpeg}$). Numbers are computed over $100$ random ImageNet images. $\epsilon_V = \frac{16}{255}$. Standard deviations are reported.}\label{tab:anomaly-detection-imagebind}
 \centering
 \fontsize{9.5}{12}\selectfont
 \begin{tabular}{laaabbb}
 \toprule
   \multirow{2.2}{*}{Augmentation method}  & \multicolumn{3}{c}{ImageBind} & \multicolumn{3}{c}{AudioCLIP} \\ 
   \cmidrule(r){2-4} \cmidrule(r){5-7}
   & \multicolumn{1}{c}{$x$} & \multicolumn{1}{c}{$x_\delta$} & \multicolumn{1}{c}{$x_{jpeg}$}  & \multicolumn{1}{c}{$x$} & \multicolumn{1}{c}{$x_\delta$} & \multicolumn{1}{c}{$x_{jpeg}$} \\ 
   \midrule
   JPEG() & $0.8784 \pm 0.04$ & $0.2257 \pm 0.07$ & $0.7303 \pm 0.08$ & $0.8512 \pm 0.06$ & $0.2149 \pm 0.08$ & $0.6198 \pm 0.09$ \\
   GaussianBlur() & $0.7493 \pm 0.09$ & $0.2108 \pm 0.07$ & $0.6447 \pm 0.09$ & $0.8068 \pm 0.08$ & $0.2070 \pm 0.07$ & $0.5637 \pm 0.09$\\
   RandomAffine() & $0.8698 \pm 0.06$ & $0.2557 \pm 0.10$ & $0.7428 \pm 0.09$ & $0.8756 \pm 0.06$ & $0.2124 \pm 0.11$ & $0.6175 \pm 0.10$ \\
   ColorJitter() & $0.8903 \pm 0.07$ & $0.4987 \pm 0.19$ & $0.8245 \pm 0.10$ & $0.9197 \pm 0.05$ & $0.4248 \pm 0.19$ & $0.7486 \pm 0.12$\\
   RandomHorizontalFlip() & $0.9713 \pm 0.05$ & $0.4006 \pm 0.09$ & $0.8670 \pm 0.07$ & $0.9688 \pm 0.02$ & $0.2949 \pm 0.07$ & $0.7070 \pm 0.07$\\
   RandomPerspective() & $0.9377 \pm 0.07$ & $0.6437 \pm 0.38$ & $0.8739 \pm 0.15$ & $0.9398 \pm 0.07$ & $0.6161 \pm 0.41$ & $0.8105 \pm 0.21$ \\
   \midrule
   Average & $ 0.8828 \pm 0.07 $ & $0.3725 \pm 0.14$ & $0.7805 \pm 0.09$ & $0.8937 \pm 0.06$ & $0.3284 \pm 0.15$ & $0.6779 \pm 0.11$\\
 \bottomrule
\end{tabular}
\end{table*}

\subsection{Anomaly Detection}

Semantically similar inputs should be embedded into similar representations.  Therefore, a plausible defense is to compare the embedding of an input with the embeddings of very similar inputs, {\em e.g.}, various augmentations. Table~\ref{tab:anomaly-detection-imagebind} shows that this defense is effective: average alignment between unmodified images and their augmentations is $0.8828$, significantly higher than $0.3725$ average alignment between the corresponding adversarial illusions and their augmentations.

The evasion attack from Section~\ref{sec:jpeg-defense} defeats this defense, too: observe that the values in the $x_{jpeg}$ column overlap with those in the $x$ column.  This illustrates the limitations of anomaly detection methods against defense-aware adversaries~\cite{sitawarin2022demystifying}.
Anomaly detection using high-dimensional statistics or enforcing a specific geometry for each modality are interesting topics for future work.

\begin{figure}[!t]
  \centering  \includegraphics[width=1.0\linewidth]{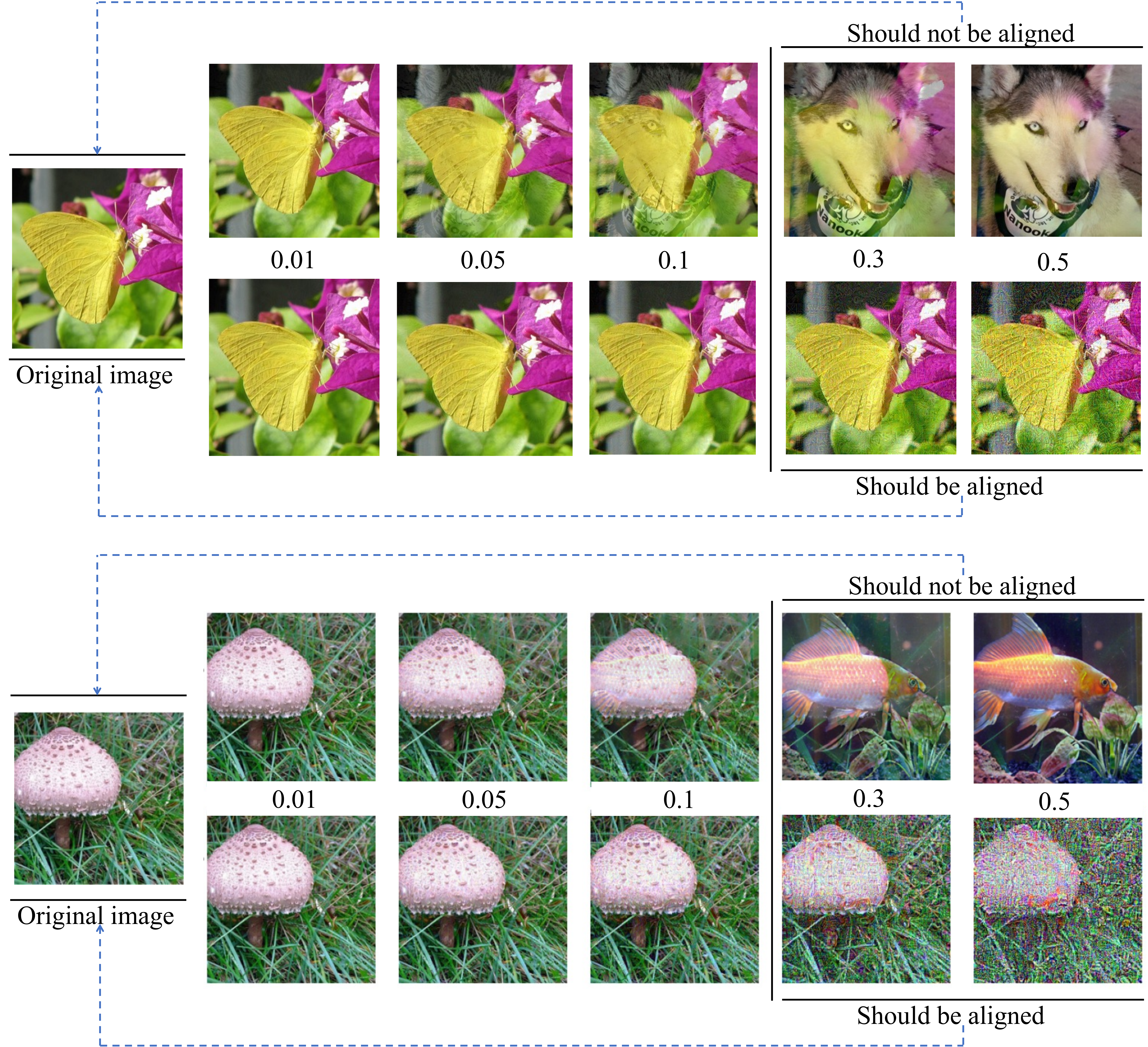}
  \vspace{-2ex}
  \caption{\textbf{Perturbations of the same magnitude can produce semantically different images or adversarial illusions.}}
  \label{fig:certification}
\end{figure}

\subsection{Certified Robustness}

\paragraphbe{Robustness to perturbations.}
Another plausible defense is to somehow train multi-modal encoders so that they are robust in the following sense: if two inputs are within $\delta$ of each other, their respective embeddings should be within $\gamma$ (using some suitable metrics for $\delta$ and $\gamma$ over the input space and embedding space, respectively).

Certified robustness~\cite{raghunathan2018certified, gowal2018effectiveness} ensures that any input perturbation within a certain bound has only a limited impact on the output of the model.  It is defined for classification tasks and cannot be directly applied to multi-modal encoders in a downstream task-agnostic way.  Given a specific downstream task, it may be possible to perform interval-bound propagation~\cite{gowal2018effectiveness} over multi-modal encoders.

Unfortunately, it is unclear which metric to use for $\delta$.  For many plausible metrics, there is a fundamental tradeoff between robustness and invariance~\cite{tramer2020fundamental}.  
If $\delta$ is too small, the adversary can use perturbations bigger than $\delta$ to evade the defense (\emph{i.e.}, the model is insufficiently robust).  If $\delta$ is too big, semantically unrelated inputs will be mistakenly aligned (\emph{i.e.}, the model is too invariant).

This observation was made for $l_p$ bounds and image classifiers in~\cite{tramer2020fundamental}.  We extend it to $l_{\infty}$ bounds, showing (by counterexample) that no $\delta$ is sufficiently robust \emph{and} does not mistakenly align semantically different inputs.

Figure~\ref{fig:certification} shows that $\delta$ must be greater than $0.3$ because there exist perturbations bigger than $0.3$ that produce a visually similar image, which should be aligned with the original.  This figure also shows that $\delta$ must be less than $0.3$ because there exist perturbations smaller than $0.3$ that produce a visually different image, which should \emph{not} be aligned with the original.  Thus, there is no ``right'' $l_{\infty}$ value for $\delta$.

\paragraphbe{Robustness to semantics-preserving transformations.}
Ultimately, embeddings should be trained so that \emph{semantically} close inputs, within and across modalities, are encoded into similar vectors in the embedding space.  There is prior work on adversarially robust ``perceptual similarity'' metrics~\cite{ghazanfari2023r}.  Perceptual similarity only captures semantic similarity insofar as images belong to the same class, \emph{i.e.}, it only aligns images with the labels of a specific image classifier.   Alignment in multi-modal embeddings is supposed to cover much broader semantic similarity.


For text, there exist semantics-preserving transformations (\emph{e.g.}, substituting words with synonyms), and models can be trained to be robust with respect to these transformations~\cite{jia-etal-2019-certified,miyato2017adversarial, yoo-qi-2021-towards-improving}.  Our attacks exploit perturbations in images and audio, where semantic similarity is not as well defined as in text.  In the image domain, robustness under different distance metrics results in vastly different performance~\cite{tramer2019adversarial}. 
We are not aware of any metric on the input
space that corresponds to task-agnostic semantic similarity, which is required to achieve certified robustness for embeddings. 

Furthermore, to ensure that the embeddings of semantically similar inputs are close to each other, the range of encoders for different modalities should overlap with very fine-grained control over changes in the embedding given a change in the input.  With such control, it is unclear how to prevent an adversary from crafting inputs closer to the target than organically aligned inputs, considering that organic alignment in the existing embeddings is relatively weak (see Section~\ref{sec:white-box-res}).

In practice, it does not appear that any existing multi-modal embedding uses certified
robustness.

\subsection{Adversarial Training}

When training classifiers, one way to ensure that small perturbations do not affect the model's output is to add perturbed images with correct class labels to the training data~\cite{madry2018towards, shafahi2019adversarial}.  In contrastive learning, adversarial training can use perturbed
images as positive samples~\cite{yu2022adversarial, kim2020adversarial, zhou2023downstream}.  The efficacy of this defense depends on the specific downstream task because some tasks (\emph{e.g.}, fine-grained classification) may require distinguishing representations that differ less than $\gamma$ (maximum embedding distance between an input and its perturbations).  Adversarial contrastive training negatively impacts relatively coarse downstream tasks such as CIFAR-10 and -100~\cite{yu2022adversarial}.  In multi-modal contrastive learning, the data are sparse, and some modalities are not directly aligned. Thus, we expect that adversarially trained embeddings significantly degrade the accuracy of downstream tasks.

\vspace{2.5ex}
\section{Conclusions}

We demonstrated that aligned multi-modal embeddings are highly vulnerable to adversarial perturbations that create \emph{cross-modal illusions}, {\em i.e.}, inputs in one modality that are aligned with semantically unrelated inputs in another modality.  The attack works across both natural and emergent alignment and is task-agnostic: adversarial alignment fools even downstream tasks of which the adversary is not aware when generating adversarial inputs.  Furthermore, it enables the adversary to use alignment targets in modalities that are not accepted by downstream applications.

We evaluated the attack for several multi-modal embeddings, including open-source embeddings such as ImageBind and AudioCLIP and proprietary, black-box embeddings such as Amazon's Titan.  We showed that the attack successfully misleads downstream tasks based on multi-modal embeddings, including audio retrieval, zero-shot classification of images, thermal images, and audio, and generation of text and images.  Finally, we discussed potential countermeasures and evasion attacks.

\vspace{2ex}
\section*{Ethical Considerations}

The only purpose of analyzing the vulnerability of multi-modal embeddings to adversarial inputs is to help develop more robust embeddings and motivate research on defenses.

\vspace{2ex}
\section*{Acknowledgments} 

This work was performed at Cornell Tech and partially supported by the NSF grant 1916717. 

\vspace{2ex}

\bibliographystyle{plain}
\bibliography{sample}

\end{document}